\documentclass[12pt]{article}

\usepackage{amsfonts,ulem}
\usepackage{amsmath}
\usepackage{amssymb,authblk}
\usepackage{amsthm}
\usepackage{array}
\usepackage{graphicx}
\usepackage[margin=1in]{geometry}
\usepackage{float}
\usepackage{booktabs}

\usepackage{mathtools}
\usepackage{psfrag}
\usepackage{mathrsfs}
 
\newcommand{\rmd}{\mathrm{d}} 
\newcommand{\rme}{\mathrm{e}}

\newcommand{\rmi}{\mathrm{i}}

\newcommand{\rmm}{\mathrm{m}}

\newcommand{\rmp}{\mathrm{p}}

\newcommand{\rms}{\mathrm{s}}

\newcommand{\rmz}{\mathrm{z}}  

\newcommand{\rmF}{\mathrm{F}}

\newcommand{\rmL}{\mathrm{L}}

\newcommand{\rmP}{\mathrm{P}}

\newcommand{\rmU}{\mathrm{U}}  
  
\newcommand{\rmW}{\mathrm{W}}

\title{\vspace{-20mm} On the Wiener--Hopf solution of water-wave interaction with a submerged elastic or poroelastic plate}

\author[1]{M. J. A. Smith}
\author[2]{M. A. Peter}
\author[3]{I. D. Abrahams} 
\author[4]{\\M. H. Meylan}

\affil[1]{\small Department of Applied Mathematics and Theoretical Physics, University of Cambridge,
Wilberforce Road, CB3 0WA, UK}
\affil[2]{Institute of Mathematics, University of Augsburg, 86135 Augsburg, Germany and Augsburg Centre for Innovative Technologies, University of Augsburg, 86135 Augsburg, Germany}
\affil[3]{Isaac Newton Institute, University of Cambridge, 20 Clarkson Road, Cambridge CB3 0EH, UK}
\affil[4]{School of Mathematical and Physical Sciences, The University of Newcastle, Callaghan, NSW 2308, Australia}

\date{}

\makeatletter
\DeclareMathOperator{\sint}{
\mathchoice%
    {\ooalign{$\displaystyle\int$\cr\large$\mkern-1mu\smile$\cr $\displaystyle\int$}}
    {\ooalign{$\int$\cr\scriptsize$\mkern3mu\smile$\cr $\int$}}
    {\ooalign{$\scriptstyle\int$\cr\tiny$\mkern2mu\smile$\cr $\scriptstyle\int$}}
    {\ooalign{$\scriptscriptstyle\int$\cr\tiny$\smile$\cr $\scriptscriptstyle\int$}}
}
\makeatother

\makeatletter
\DeclareMathOperator{\fint}{
\mathchoice%
    {\ooalign{$\displaystyle\int$\cr\large$\mkern-1mu\frown$\cr $\displaystyle\int$}}
    {\ooalign{$\int$\cr\scriptsize$\mkern3mu\frown$\cr $\int$}}
    {\ooalign{$\scriptstyle\int$\cr\tiny$\mkern2mu\frown$\cr $\scriptstyle\int$}}
    {\ooalign{$\scriptscriptstyle\int$\cr\tiny$\frown$\cr $\scriptscriptstyle\int$}}
}
\makeatother

\begin{document}
\maketitle
\vspace{-110mm}
\begin{center}
\small Accepted Manuscript
\end{center}	
\vspace{90mm}
\begin{abstract}
    A solution to the problem of water-wave scattering by a semi-infinite submerged thin elastic plate, which is either porous or non-porous, is presented using the Wiener--Hopf technique.  The derivation of the Wiener--Hopf equation is rather different from that which is used traditionally in water-waves problems, and it leads to the required equations directly.  It is also shown how the solution can be computed straightforwardly using Cauchy-type integrals, which avoids the need to find the roots of the highly non-trivial dispersion equations.  We illustrate the method with some numerical computations, focusing on the evolution of an incident wave pulse which illustrates the existence of two transmitted waves in the submerged plate system. The effect of the porosity is studied, and it is shown to influence the shorter-wavelength pulse much more strongly than the longer-wavelength pulse. 
\end{abstract}

\section{Introduction}
The field of wave--structure interactions is concerned with the propagation of wave energy within a fluid (liquid or gas) and the coupled mechanical response excited in an accompanying body. Such interactions are more involved than simple scattering processes as they feature wave energy propagation in multiple forms, for example, when an airflow excites a mechanical response in an aeroplane wing (aeroelastic flutter) or when a water wave excites bending waves in a floating elastic plate  (floating ice sheets). Wave--structure interactions are studied in a range of disciplines owing to their incredibly diverse and wide-ranging engineering applications, from aircraft design to marine hydrodynamics and coastal engineering \cite{howe1998acoustics,linton2001handbook}.  The study of wave--structure interactions, particularly those involving flexible horizontal plates, has been the subject of extensive investigation over the past several decades. We refer to the review papers \cite{squire2020ocean,squire07} for a summary of the research, with particular focus on the propagation of waves in ice-infested oceans.  

Several canonical wave--structure-interaction problems exist for elastic plates, the most prominent of which is the linear scattering of waves by a floating (thin) elastic semi-infinite plate (i.e.~a floating plate of negligible submergence). The first attempt to solve this problem used the Wiener--Hopf method \cite{evans1968wave}; however, the solution was not completed in this work. The problem was finally solved using variational methods  \cite{fox_squire91} and the various papers inspired by this work are summarised by Fox \& Squire \cite{fox_squire94}. In fact, the Wiener--Hopf solution to the semi-infinite floating elastic plate problem started by Evans \& Davies \cite{evans1968wave} was eventually determined by several groups  \cite{balmforth_craster99,tkacheva01,chung_fox02}. Another canonical problem is the semi-infinite submerged elastic plate, which acts as a model for a broad class of problems in which a flexible body is immersed within a fluid. For this problem, analogous to the case of a two-layer fluid, waves exist at all interfaces: at both the free surface and along the submerged plate. In fact, for both the submerged elastic plate and two-layer fluid problems there are two transmitted waves on the free surface \cite{sahoo2012mathematical,das2018flexural}; for the submerged plate problem, the first transmitted wave relates straightforwardly to the incident field and the second transmitted field emerges due to fluid--structure coupling with the elastic plate. As discussed below, we demonstrate here that the presence of poroelasticity in the submerged plate can strongly suppress one of the transmitted waves (the plate-interaction wave), leaving only one significant transmitted surface wave that attenuates slowly. The solution for the semi-infinite submerged elastic plate was found first using the eigenfunction matching method \cite{ulhassan2009water} and later by the Wiener--Hopf method \cite{williams2012wiener}. In the water-wave context, horizontal submerged plates are particularly popular to dissipate water-wave energy, e.g.~in breakwaters or at the end of wave flumes, as they have little effect on horizontal currents \cite{Wang1999}. 

 The porous plate is a natural and significant extension to the plate problems outlined above and has been the subject of recent attention in a range of applications \cite{jaworski2013aerodynamic,zheng2020wave}. In particular, it naturally dissipates energy. The problem of a submerged semi-infinite porous rigid dock was solved by the Wiener--Hopf method in Evans \& Peter  \cite{evans2011asymptotic}. 
Since porosity is often associated with thin objects (compared to its other dimensions) the porous plate  responds elastically if subjected to incident waves unless they are very short in wavelength. On the other hand, if horizontal (elastic) plates are to be used to dissipate wave energy, it makes sense to make them porous. Therefore, it is natural to consider the water-wave interaction with a porous elastic plate. In fact, porous plates have been studied lately with the solution found by eigenfunction matching   \cite{behera2015hydroelastic}. We also refer to Meylan et al.~\cite{porousplate} for a recent review of the literature on floating or submerged porous plates. 
The interest in porous plates extends beyond the water-waves community; for example, there is considerable interest within the aeroacoustics community on efficiently computing the response of porous plates to high-frequency air flows \cite[\S 5.4.2]{howe1996influence},\cite{jaworski2013aerodynamic}, as such systems are leading-order models for investigating noise-emission suppression by aircraft. There is also a considerable body of work from within the acoustics community \cite{cannell1975edge,crighton1991fluid,zhang1995radiation,david2000existence}.
 
As identified above, the Wiener--Hopf method is a robust solution procedure that is widely used in many fields \cite{noble1959methods,howe1998acoustics,linton2001handbook,lawrie2007brief}. It provides a simple semi-analytical treatment for problems which are typically solved by numerical techniques otherwise. The method relies on domain decomposition, i.e.~splitting a function analytic on the real line into functions analytic in the upper and lower half-planes. In the standard application of the method, this decomposition is accomplished by calculating and sorting the zeros of a dispersion equation or equivalent  \cite{balmforth_craster99,linton2001handbook,tkacheva01,chung_fox02,evans2011asymptotic,williams2012wiener}. Often, this is numerically challenging as there is no systematic method to find the zeros. An alternative approach is to compute this splitting via a Cauchy-type integral, which avoids this need of factorisation \cite[\S 1.3]{noble1959methods}. For the submerged porous dock, this was first achieved by  Evans \& Peter \cite{evans2011asymptotic}. 
 
In this paper, we present a Wiener--Hopf solution to the problem of a semi-infinite poroelastic plate submerged in an incompressible fluid of finite depth. Our derivation does not exploit the factorisation of the dispersion equations but instead uses Cauchy integrals. We find that the presence of porosity can impact the transmitted surface-wave behaviour of the fluid significantly, forcefully suppressing the secondary short-wavelength wave and attenuating the long-wavelength wave (corresponding to the incident field). This finding may have a bearing on the development of structures for coastal-engineering applications. 

The outline of the paper is as follows. In Section \ref{sec:gov_eq}, we present the governing equations for the fluid--structure interaction (coupling a potential flow to a submerged, lossless Kirchhoff--Love plate), giving rise to an unknown polynomial $P=P(s)$ which emerges from repeated integration by parts. In Section \ref{sec:prod_decomp}, we present the product decomposition of the Wiener--Hopf kernel involving Cauchy-type integrals and determine the asymptotic behaviour of the system at infinity, giving rise to an unknown polynomial $J=J(s)$. In Section \ref{sec:dtpj}, we determine these two polynomials using the plate-edge boundary conditions and by imposing appropriate analyticity conditions in the upper- and lower-half planes. Having solved the Wiener--Hopf system, we then construct the total potential using residue calculus in Section  \ref{sec:solrep}, outlining the energy-balance relation for the system and presenting numerical results. In Section \ref{sec:extporo}, we consider the extension of the system to submerged poroelastic plates, incorporating the effect of a porous flow across the plate (following Darcy's law), which follows straightforwardly from the non-porous case thanks to our way of deriving the Wiener--Hopf equation. We also produce comparative numerical results to the lossless case. Finally, concluding remarks are given in Section \ref{sec:concl_rem}.
\section{Governing equations} \label{sec:gov_eq}
In this work, we consider the problem of wave propagation through a waveguide comprising a horizontal semi-infinite elastic plate  submerged within a fluid domain, which possesses  a rigid horizontal sea floor and free fluid surface. The fluid medium is governed by   the three-dimensional Laplace equation
\begin{equation}
\label{eq:govfluideq}
\Delta \Phi(x,y,z;t) = 0, \quad \mbox{for } (x,y,z) \in \Omega,
\end{equation}
where $\Delta = \partial_x^2 +\partial_z^2 +\partial_z^2 $, $\Omega = \left\{(x,y) \right. \in \mathbb{R}_2 \times z \in \left.(-h,0)\right\}\backslash \Gamma_\rmp$ is the fluid domain, $\Gamma_\rmp =  \left\{x>0 \times y \in \mathbb{R} \times  z=-d \right\}$ is the elastic plate domain, where $d<h$, and $\Phi$ is the three-dimensional fluid velocity potential. Note that the operator in \eqref{eq:govfluideq} above is an appropriate model for fluid flows that are both incompressible and irrotational (i.e., where the fluid density is constant and so the conservation of mass condition is given in terms of the fluid velocity $(u_x,u_y,u_z)$ alone as $\partial_x u_x + \partial_y u_y + \partial_z u_z = 0$ and the flow velocity is decomposed as $(u_x,u_y,u_z) = (\partial_x \Phi,\partial_y \Phi,\partial_z \Phi)$, respectively). The bending response of the submerged uniform plate is given by the Kirchhoff--Love thin plate equation
\begin{equation}
\label{eq:kirchofflove}
D \Delta_\parallel^2 w_b(x,y;t) +    h_\rmP \rho  \partial_t^2 w_b(x,y;t)= -q, \quad \mbox{for } (x,y) \in \Gamma_\rmp,
\end{equation}
where  $D =   h_\rmP^3 E/(12[1-\nu^2])$ is the bending stiffness, $h_\rmP$ is the total plate thickness, $E$ is the Young's modulus, $\nu$ is the Poisson ratio, $\Delta_\parallel = \partial_x^2 + \partial_y^2$ is a two-dimensional Laplace operator, $w_b(x,y;t)$ is the out-of-plane displacement, $\rho$ is the plate mass density, and $q$ is the plate loading.

 We   consider incident potential fields of the form $\Phi_\mathrm{inc}(x,y,z)= \phi_\mathrm{inc}(x,z)\exp(\rmi k_y^\mathrm{inc} y - \rmi \omega t)$, where $(k_x^\mathrm{inc},k_y^\mathrm{inc},k_z^\mathrm{inc})$ denotes the incident wavevector, and $\omega$ is the angular frequency, in the long-wavelength limit   $k_y^\mathrm{inc} h\rightarrow 0$. That said, the formulation outlined in-text is readily extended to consider skew incidence. Accordingly we decompose all  potentials in the analogous form  $\Phi(x,y,z;t)= \phi(x,z)\exp(\rmi k_y^\mathrm{inc} y - \rmi \omega t)$, decompose the plate displacement as $w_b(x,y;t) = w(x,y)\exp(-\rmi \omega t)$, and consider	 a problem independent of $y$ (i.e., where the $y$ dependence in $\Phi$ is the same as in $\Phi_\mathrm{inc}$).  
 
 Furthermore, for the plate forcing we assume a linearised Bernoulli response for the pressure on the submerged beam
\begin{equation}\label{eq:forcingq}
q = P(x,y,-d)\big|_{-}^{+} = \left\{ \rho_\rmF  \,g   d - \rho_\rmF \, \partial_t \Phi(x,y,-d;t) \right\}\big|_{-}^{+} = \rmi \omega \rho_\rmF \phi \big|_{-}^{+} \exp(-\rmi \omega t),
\end{equation}
where the $\rho_\rmF  \,g   d$ term describes the hydrostatic pressure of the fluid at depth $z=-d$, and the second term denotes the pressure contribution from  the fluid motion. 
 Here we define $\phi\big|_{-}^{+}=\phi(x,-d_+) - \phi(x,-d_-)$ as the jump discontinuity in the potential across the thin plate, where $-d_+$ is the upper side and $-d_-$   the lower side of the plate, $\rho_\rmF$ is the density of the fluid,  and $g = 9.8 \, \rmm  \cdot\rms^{-2}$ is the gravitational acceleration constant.

   \begin{figure}[t]
       \includegraphics[width=0.45\textwidth]{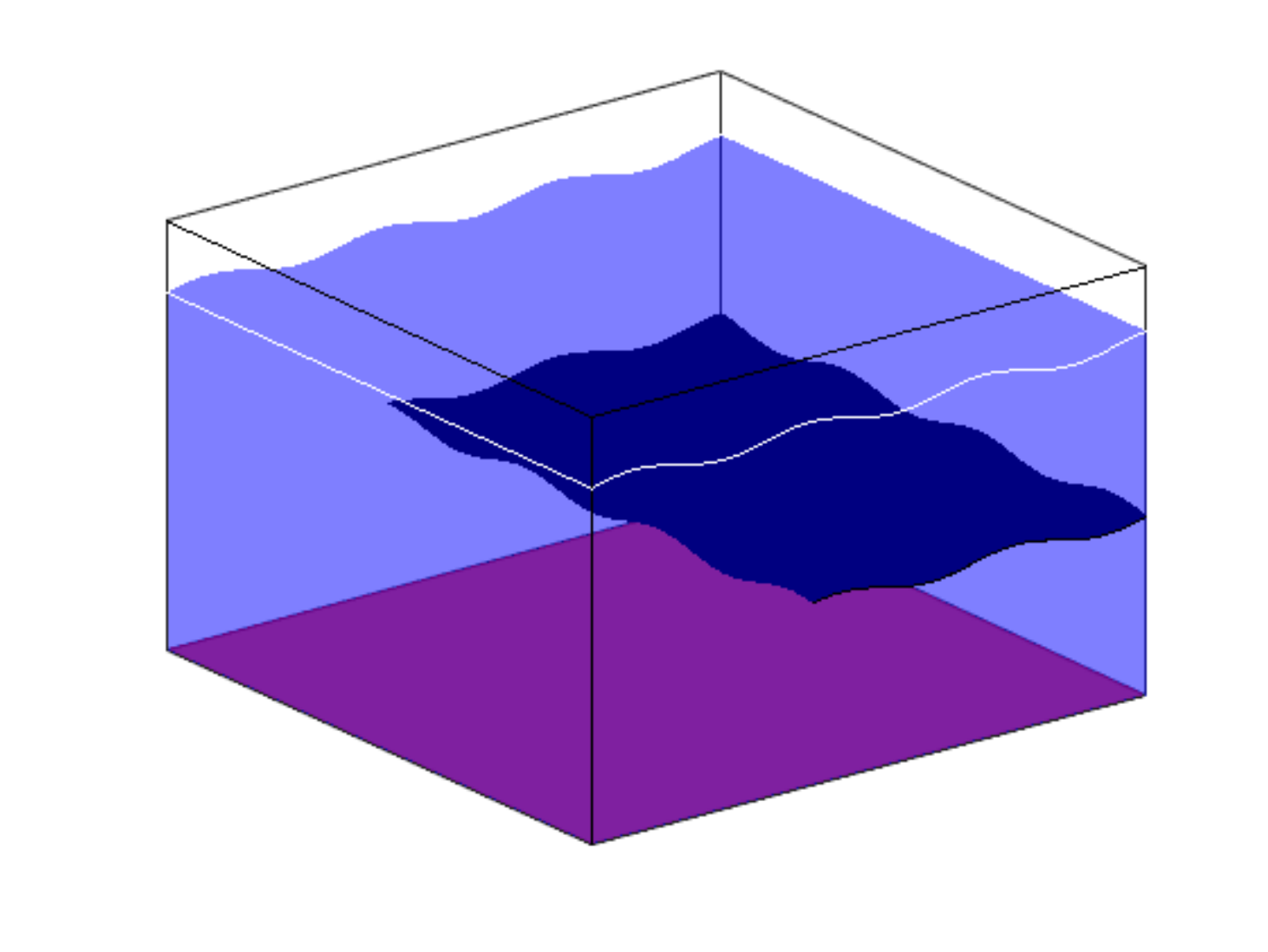}
     \hfill
       \includegraphics[width=0.45\textwidth]{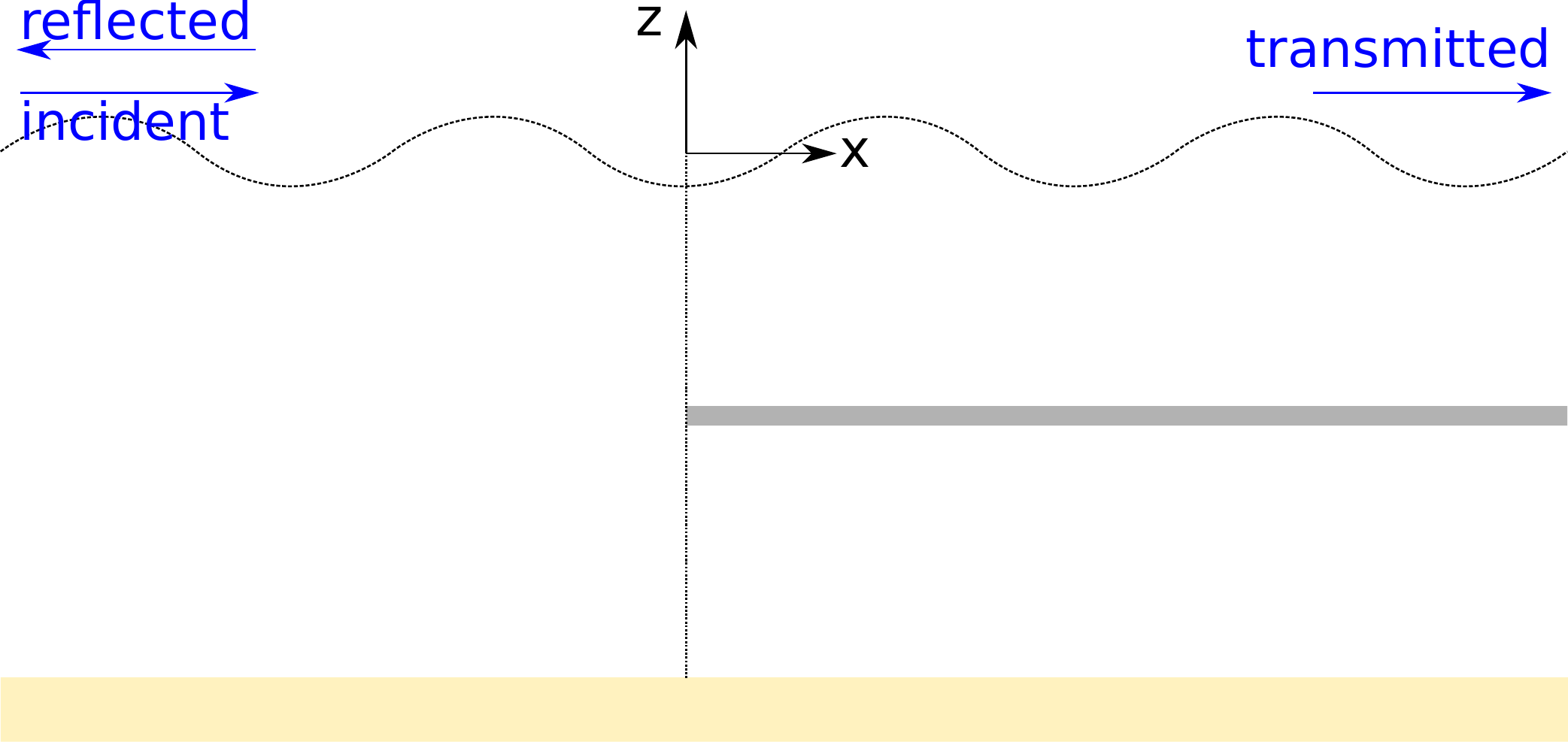}
     \caption{\small  (a) Three-dimensional segment, and (b) two-dimensional cross section of a submerged semi-infinite elastic plate at depth $z=-d$ with a rigid sea floor at depth $z=-h$ and a free surface at $z=0$. Incident surface waves excite reflected and transmitted fields travelling to the left and right, respectively. \label{fig:schem}}
 
   \end{figure}

Finally, we impose a kinematic condition between the plate and fluid requiring that the normal velocity of the fluid matches the normal velocity of the plate
\begin{equation}\label{eq:kin_elpl}
\partial_t w_b(x,y;t) =  \partial_z\Phi(x,y,-d;t),\quad \mbox{for } (x,y) \in \Gamma_\rmp,
\end{equation}
which implies continuity of the normal fluid velocity across the plate.  Accordingly, we arrive at the two-dimensional  system of    equations considered in this work involving a single scalar dependent variable $\phi(x,z)$ satisfying Laplace's equation 
\begin{align}
\label{eq:2dlaplacefluid}
\Delta_\perp \phi(x,z) = 0, \quad \mbox{for } (x,z) \in \Omega,
\end{align}
where $\Delta_\perp = \partial_x^2 + \partial_z^2$ is a two-dimensional Laplace operator, with the free surface, elastic plate, fluid velocity, and rigid sea floor conditions
\begin{subequations}
\begin{align}
\label{eq:freesurf1}
\partial_z \phi(x,z) - \alpha \phi(x,z)&= 0, \quad \mbox{for } z=0\phantom{-} \mbox{ and }x \in \mathbb{R}, \\
\label{eq:plate1}
(\partial_x^4 - \mu^4)\partial_z\phi(x,z) + \beta \phi(x,z)\big|_{-}^{+}&= 0, \quad \mbox{for } z=-d \mbox{ and }x>0, \\
\label{eq:fluidveljump1}
\partial_z \phi(x,z)\big|_{-}^{+} &= 0, \quad \mbox{for } z=-d\mbox{ and }x \in \mathbb{R}, \\
\label{eq:rigidfloor1}
\partial_z \phi(x,z) &= 0, \quad \mbox{for } z=-h\mbox{ and }x \in \mathbb{R}, 
\end{align}
\end{subequations}
where $\alpha = \omega^2/g$,   $\mu^4 = \rho h_\rmP \omega^2/D  $, and $\beta = \rho_\rmF   \omega^2/D $. The conditions at the plate edge (i.e., clamped, free edge, or simply supported conditions) are specified later. For the pressure jump across the plate, the leading order asymptotic behaviour is easily shown to be
\begin{equation}
\label{eq:asyjump}
\lim_{x \rightarrow 0^-} \phi(x,-d)\big|_{-}^{+} \sim  O(x^{1/2}), 
\end{equation}
near the plate edge, regardless of the conditions, and that outgoing wave (radiation)  conditions are satisfied at infinity \cite[Eq.~(1.29)]{linton2001handbook}.
Note that in place of the pressure condition \eqref{eq:asyjump} we could equivalently impose a condition on the flow velocity \cite{evans2011asymptotic}
\begin{equation}
\label{eq:asyfluidflow} 
\lim_{x \rightarrow 0^-}   \sqrt{\left[\partial_x \phi(x,-d)\right]^2+\left[\partial_z \phi(x,-d)\right]^2} \sim O(x^{-1/2}),
\end{equation}
as these are equivalent from a conservation of momentum argument. Both forms are presented here for later convenience.  Next, we decompose the field in terms of an incident and scattered field
\begin{equation}
\label{eq:decomp}
\mathcal{A} \phi(x,z) =   \phi^\mathrm{inc}(x,z)+ \phi^\rms(x,z),
\end{equation}
and examine the possible forms of the incident potential for this system in detail. Note that we introduce the arbitrary scaling factor $\mathcal{A}$ for later convenience (for example, if normalising the potential to have unit amplitude in displacement). Up until Section \ref{sec:solrep}, the scaling $\mathcal{A}=1$ is taken without loss of generality. The scattered potential $\phi^\rms$ is subject to radiation conditions that the waves are outgoing at infinity and we impose these as part of the solution procedure.

\subsection{Incident potential}
The incident potential, which we take to come from $x=-\infty$, must satisfy both the free-surface \eqref{eq:freesurf1} and rigid-floor \eqref{eq:rigidfloor1} conditions, irrespective of the presence of the submerged plate. Accordingly, we propose the modulated plane-wave ansatz
\begin{equation}
\label{eq:pwansantz}
\phi^\mathrm{inc}(x,z)= F \rme^{\rmi \ell x} \cosh(\ell(z+h)),   
\end{equation}
where $F$ is arbitrary. The above ansatz satisfies the rigid-floor condition by inspection, and admits the dispersion equation
\begin{equation}
\label{eq:freesurfdispeq}
\ell \tanh(\ell h) = \alpha,
\end{equation}
from the free-surface condition. This  dispersion equation is symmetric in $\ell$ and so has   one real positive root and one real negative root. We choose the value of $\ell$ to be the positive real solution of this equation, which corresponds to waves travelling to the right. By substituting the decomposition \eqref{eq:decomp} and plane-wave ansatz \eqref{eq:pwansantz} into the plate condition \eqref{eq:plate1} we obtain
\begin{equation}
\label{eq:plateforcingincF}
(\partial_x^4 - \mu^4)\partial_z\phi^\rms(x,z) + \beta \phi^\rms(x,z)\big|_{-}^{+}= - \ell(\ell^4 - \mu^4)  \sinh(\ell c) F \rme^{\rmi \ell x},
\end{equation}
where $c=h-d$, and so we specify the incident potential
\begin{equation}
\label{eq:pwansantz2}
\phi^\mathrm{inc}(x,z)= \frac{-\cosh(\ell(z+h))}{\ell (\ell^4 - \mu^4)  \sinh(\ell c) } \rme^{\rmi \ell x} ,   
\end{equation}
for a line source placed at $x=-\infty$ in our free-surface and rigid-floor domain. This choice of $F$ ensures a simple form for the plate forcing, i.e., the right-hand side of \eqref{eq:plateforcingincF}. 

\subsection{Scattered potential}
From the decomposition \eqref{eq:decomp} above, the system for the scattered potential takes the form
\begin{align}
\label{eq:laplace2}
\Delta_\perp \phi^\rms(x,z) = 0, \quad \mbox{for } (x,z) \in \Omega,
\end{align}
with the updated boundary conditions
\begin{subequations}
\begin{align}
\label{eq:freesurf2}
\partial_z \phi^\rms(x,z) - \alpha \phi^\rms(x,z)&= 0, \quad \mbox{for } z=0\phantom{-} \mbox{ and }x \in \mathbb{R}, \\
\label{eq:plate2}
(\partial_x^4 - \mu^4)\partial_z\phi^\rms(x,z) + \beta \phi^\rms(x,z)\big|_{-}^{+}&=  \rme^{\rmi \ell x}, \quad \mbox{for } z=-d \mbox{ and }x>0, \\
\label{eq:fluidveljump2}
\partial_z \phi^\rms(x,z)\big|_{-}^{+} &= 0, \quad \mbox{for } z=-d\mbox{ and }x \in \mathbb{R}, \\
\label{eq:rigidfloor2}
\partial_z \phi^\rms(x,z) &= 0, \quad \mbox{for } z=-h\mbox{ and }x \in \mathbb{R}. 
\end{align}
\end{subequations}
Next we  introduce the Fourier transform 
\begin{equation}
\label{eq:fouriertrans} 
\Psi(s,z) = \int_{-\infty}^{\infty} \phi^\rms(x,z)\, \rme^{\rmi s x}\, \rmd x, 
\end{equation}
and apply this to the two-dimensional Laplace equation \eqref{eq:laplace2} to obtain
\begin{equation}
( -s^2 + \partial_z^2)\Psi(s,z)  = 0,
\end{equation}
which has the general solution
$
\Psi(s,z)  = G(s) \cosh(sz) + H(s) \sinh(sz).
$
We then   introduce a (fictitious) partition of the domain in the $z$ direction  as
\begin{align}
\label{eq:psipartition}
\Psi(s,z) = \begin{cases} 
      \Psi^\rmU(s,z)  &  \mbox{for }  -d<z<0, \\
       \Psi^\rmL(s,z)  & \mbox{for }  -h<z<-d.  
   \end{cases}
\end{align}
Consequently for  $-d<z<0 $ we solve
\begin{subequations}
\begin{align}
( -s^2 + \partial_\rmz^2)\Psi^\rmU(s,z)  = 0, \quad \mbox{with } \\
 \partial_z\Psi^\rmU(s,z)- \alpha \Psi^\rmU(s,z) = 0, \quad \mbox{for } z=0,
\end{align}
\end{subequations}
admitting the form $\Psi^\rmU(s,z) = G^\rmU(s) \left[  \cosh(s z) +(\alpha/s) \sinh(s z) \right]$. Similarly, for $-h<z<-d $ we solve
\begin{subequations}
\begin{align}
( -s^2 + \partial_\rmz^2)\Psi^\rmL(s,z)  = 0, \quad \mbox{with } \\
 \partial_z\Psi^\rmL(s,z)  = 0, \quad \mbox{for } z=-h,
\end{align}
\end{subequations}
to obtain $\Psi^\rmL(s,z) = H^\rmL(s) \cosh(s(z+h))/\sinh(sh)$. These forms are then matched on $z=-d$ by imposing continuity of the   derivative \eqref{eq:fluidveljump2}, i.e., 
\begin{equation}
\label{eq:matchcond}
\partial_z \Psi^\rmU(s,-d) = \partial_z \Psi^\rmL(s,-d),
\end{equation}
admitting the relation
\begin{equation}
\label{eq:CsvsDs}
G^\rmU(s) = \frac{ s \sinh(s c) }{ \sinh(sh) (\alpha \cosh(s d) - s \sinh(s d) ) } H^\rmL(s).
\end{equation}
At this stage the form of $H^\rmL(s)$, and therefore the complete $x$ dependence in the solution, is unknown, and it is through the Wiener--Hopf technique that this dependence is found.
\subsection{Field decomposition}
Having determined the form of the fluid eigenfunctions, we now introduce the decomposition  
\begin{equation}
\label{eq:decompphijump}
\phi^\rms(x,-d)\big|_-^+=   
\begin{cases} 
     0 & \mbox{for }  x<0, \\
      a(x)& \mbox{for } x>0,  
   \end{cases}
\end{equation}
where  $a(x)$ is unknown and for $x<0$   there is no pressure discontinuity    in the fluid domain. Similarly, we write the normal velocity as 
\begin{equation}
\label{eq:phizbx}
 \partial_z \phi^\rms(x,-d) = 
\begin{cases} 
     b(x) & \mbox{for }  x<0, \\
      m(x)& \mbox{for } x>0,  
   \end{cases}
\end{equation}
where $b(x)$ and $m(x)$ are unknown. Evaluating the Fourier transform of  $\phi^\rms(x,-d)\big|_-^+$ we obtain
\begin{align}
\label{eq:fouriertransApD}
\Psi(s,-d)\big|_-^+ &=  \int_{0}^{\infty} a(x)\, \rme^{\rmi s x}\, \rmd x = A^+(s) = \left[\frac{\sinh(sc) \xi(s) - \cosh(sc) \eta(s)}{ \sinh(sh) \eta(s) }\right] H^\rmL(s),
\end{align}
where    $A^+(s)$ is yet unknown, and
\begin{subequations}
\begin{align}
\xi(s)   &=s \cosh(s d) -  \alpha \sinh(s d), \\
\eta(s)  &=\alpha \cosh(s d) -  s \sinh(s d) 
\end{align}
\end{subequations}
from the matched eigenfunction expansions earlier, with superscript plus ($+$) notation denoting analyticity in the upper-half plane.
Next, we consider the Fourier transform of the plate operator at $z=-d$ for all $x \in \mathbb{R}$, which admits
\begin{multline}
\label{eq:fourierplateallx}
\int_{-\infty}^{\infty} (\partial_x^4 - \mu^4)\partial_z\phi^\rms(x,-d)  \, \rme^{\rmi s x}\rmd x \\
= \int_{-\infty}^{0} (\partial_x^4 - \mu^4)\partial_z\phi^\rms(x,-d)  \, \rme^{\rmi s x} \rmd x + \int_{0}^{\infty} (\partial_x^4 - \mu^4)\partial_z\phi^\rms(x,-d)  \,\rme^{\rmi s x} \rmd x.
\end{multline}
The   left-hand side of \eqref{eq:fourierplateallx}, after considering the Fourier transform definition   \eqref{eq:fouriertrans} and jump condition  \eqref{eq:fouriertransApD} above, takes the form
\begin{subequations}
\begin{multline}
\label{eq:lhsplateop}
\int_{-\infty}^{\infty} (\partial_x^4 - \mu^4)\partial_z\phi^\rms(x,-d)  \, \rme^{\rmi s x} \rmd x = (s^4 - \mu^4)\partial_z\Psi(s,-d)  \\
= \left[\frac{ s (s^4-\mu^4)\sinh(sc) \eta(s) }{\sinh(sc) \xi(s) - \cosh(sc) \eta(s)}\right]A^+(s).
\end{multline}
The first term on the right-hand side of \eqref{eq:fourierplateallx}, after repeated integration by parts, admits
\begin{multline}
\label{eq:intbyparts4x}
 \int_{-\infty}^{0} (\partial_x^4 - \mu^4)\partial_z\phi^\rms(x,-d)  \, \rme^{\rmi s x} \rmd x   \\
 =  \left[ (\rmi   \partial_z \phi^\rms_0 )s^3  - (\partial_x \partial_z \phi^\rms_0 )  s^2  - (\rmi  \partial_x^2 \partial_z \phi^\rms_0 ) s + (\partial_x^3 \partial_z \phi^\rms_0 ) \right] + (s^4-\mu^4)  \int_{-\infty}^{0} \partial_z \phi^\rms(x,-d) \, \rme^{\rmi s x} \rmd x \\
= P(s) +(s^4 - \mu^4) B^-(s),
\end{multline}
where boundary contributions from negative infinity are excluded by appropriate regularisation, the  subscript zero notation denotes   the limit
\begin{equation}
\label{eq:limcoeffdef}
 \partial_z \phi^\rms_0 = \lim_{x\rightarrow 0} \partial_z \phi^\rms(x,-d),
\end{equation}
etc., which represent constant coefficients   in the cubic polynomial $P(s)$, and 
\begin{equation}
B^-(s)  = \int_{-\infty}^{0} b(x) \rme^{\rmi s x} \, \rmd x,
\end{equation}
is as yet unknown except that it is analytic in the lower-half plane.  The second term on the right-hand side of \eqref{eq:fourierplateallx} is simplified by the boundary condition \eqref{eq:plate2} and jump decomposition \eqref{eq:decompphijump} and so we obtain
\begin{equation}
 \int_{0}^{\infty} (\partial_x^4 - \mu^4)\partial_z\phi^\rms(x,-d)  \, \rme^{\rmi s x} \rmd x = -\beta \int_{0}^{\infty}  a(x) \, \rme^{\rmi s x} \rmd x +  \int_{0}^{\infty}  \rme^{\rmi (\ell + s) x} \rmd x = -\beta A^+(s) + \frac{\rmi}{(s+\ell)_+},
\end{equation}
\end{subequations}
where the result for the improper integral $\int_{0}^{\infty}  \exp\left\{\rmi (\ell + s) x \right\} \rmd x$ follows from appropriate regularisation.
By combining all of the above, we finally arrive at the Wiener--Hopf equation
\begin{equation}
\label{eq:wheqn}
K(s) A^+(s) = P(s) +(s^4 - \mu^4) B^-(s)    + \frac{\rmi}{(s+\ell)_+},
\end{equation}
where
\begin{equation}
\label{eq:Ks}
K(s) = s (s^4 - \mu^4)   \left[\frac{ \sinh(sc) \eta(s) }{   \sinh(sc) \xi(s) - \cosh(sc) \eta(s) }\right] + \beta,
\end{equation}
and the Wiener--Hopf system \eqref{eq:wheqn} holds in an infinite strip including the real line, but indented below (above) any isolated singularities on the positive (negative) real line so that radiation conditions are satisfied. In fact, these poles come from the forcing term and the Wiener--Hopf kernel function  $K(s)$,  which is regular along the real line (suitably indented) and also possesses poles along the imaginary $s$ axis. A key step in the solution procedure is the factorisation of $K(s)$, which is the focus of the next section.

\section{Product decomposition of the kernel  K(s)} \label{sec:prod_decomp}
It follows from inspection that the denominator of  $K(s)$ is the dispersion equation   \eqref{eq:freesurfdispeq}    and thus has poles at $s = \pm \ell$. Also, we find that 
\begin{equation}
\label{eq:Ksinfasy}
\lim_{s \rightarrow \pm \infty} K(s) \sim -| s|^5
\end{equation}   and so we propose the factorisation \cite{abrahams2000application}
\begin{equation}
K(s)  = L(s) Q(s),
\end{equation}
where
\begin{subequations}
\begin{align}
L(s) &= -\frac{1}{2}s (s^4 - \mu^4) \frac{\cosh(\pi s)}{\sinh(\pi s)}, \\
 Q(s) &=   2 \tanh(\pi s)  \left[\frac{\sinh(cs) \eta(s) }{ \alpha \cosh(s h) - s \sinh(s h)   } - \frac{\beta}{s(s^4-\mu^4)}\right]  .    
\end{align}
\end{subequations}
With this decomposition, we   have that  $Q(s) \rightarrow 1$ as $ s \rightarrow \pm \infty$. Using the well-known Euler reflection identities \cite[Eqs. (6.1.30), (6.1.31)]{abramowitz1972handbook}
\begin{subequations}
\begin{align}
\sinh(\pi z) \Gamma(1+\rmi z)\Gamma(1 - \rmi z) &= \pi z,\\
\cosh(\pi z) \Gamma(\tfrac{1}{2}+\rmi z)\Gamma(\tfrac{1}{2} - \rmi z)&= \pi  ,
\end{align}
\end{subequations}
where $\Gamma(z)$ denotes the Euler Gamma function, then $L(s)$ is readily factorised    as
\begin{subequations}
\begin{align}
L^+(s) =  \frac{\rmi(s+\mu)(s+\rmi\mu)\Gamma(1 - \rmi s)}{\sqrt{2}\Gamma(\tfrac{1}{2} - \rmi s)}, \\
L^-(s) =  \frac{\rmi(s-\mu)(s-\rmi\mu)\Gamma(1 + \rmi s)}{\sqrt{2}\Gamma(\tfrac{1}{2} + \rmi s)},
\end{align}
\end{subequations}
thereby removing the need to evaluate infinite products, as is common with other representations (see for example Linton \& McIver \cite[\S 5.1.3]{linton2001handbook}). The product factorisation for $Q(s)$ comes from Cauchy's integral formula (see Noble \cite[pp. 13]{noble1959methods}) and takes the form
\begin{subequations}
\begin{align}
 Q^+(s)  &= \exp\left\{ \frac{1}{2\pi\rmi} \sint  \frac{\log(Q(z))}{z-s} \,\rmd z\right\}, \\
 Q^-(s)  &= \exp\left\{ \frac{-1}{2\pi\rmi} \fint  \frac{\log(Q(z))}{z-s} \,\rmd z \right\} ,
\end{align}
\end{subequations}
where  $\smile$ represents a path traversing beneath $z=s$, and conversely $\frown$ above $z=s$, from $-\infty$ to $\infty$. More specifically, starting from Cauchy's integral formula with a closed contour centred about the point $z=s$, where $s$ is any point of analyticity of the function, we take this point to lie  inside the region in which the Wiener--Hopf equation \eqref{eq:wheqn} is defined (an infinite strip including the real line, but indented below (above) any isolated singularities on the positive (negative) real line). This closed contour can then be extended to $z=\pm \infty$ inside this region of analyticity, which yields two infinite contours, one passing below $s$, denoted by $\smile$, and the other above, denoted by $\frown$.  

 Thus, we have that  $K(s) = K^+(s) K^-(s)$ where
\begin{equation}
  K^+(s) = L^+(s) Q^+(s)  \quad \mbox{ and } \quad  K^-(s) = L^-(s) Q^-(s)     .
\end{equation}
 In practical terms, only one of the above  need be evaluated due to the symmetries $L^+(-s) = L^-(s)$ and $Q^+(-s) = Q^-(s)$, i.e., $K^+(-s) = K^-(s)$. 
Returning to the Wiener--Hopf equation \eqref{eq:wheqn} we introduce the decomposition for $K(s)$ to obtain
\begin{subequations}
\begin{equation}
\label{eq:systemtakei}
  K^+(s) A^+(s) = \frac{ P(s)}{K^-(s)} +(s^4 - \mu^4)   \frac{B^-(s)}{K^-(s)}  + \frac{\rmi}{(s+\ell)_+}\frac{1}{K^-(s)} ,
\end{equation}
where the final term on the right-hand side of \eqref{eq:systemtakei} is a product of two functions, one regular in the upper- and the other regular in the lower-half plane. 
We now introduce the additive decomposition 
\begin{equation}
 \frac{\rmi}{(s+\ell)_+}\frac{1}{K^-(s)} =  \frac{\rmi}{(s+\ell)_+}\left(     \frac{1}{K^-(s)}   -  \frac{1}{K^-(-\ell)} \right) +  \frac{\rmi}{(s+\ell)_+} \frac{1}{K^-(-\ell)},
\end{equation}
where the first term on the left-hand side has a removable pole at  $s = -\ell$ and thus is regular. Consequently, we obtain the form
\begin{equation}
\label{eq:WHfinal}
  K^+(s) A^+(s) - \frac{\rmi}{(s+\ell)_+} \frac{1}{K^-(-\ell)}= \frac{ P(s)}{K^-(s)} +(s^4 - \mu^4)   \frac{B^-(s)}{K^-(s)}  + \frac{\rmi}{(s+\ell)_+}\left(     \frac{1}{K^-(s)}   -  \frac{1}{K^-(-\ell)} \right)   ,
\end{equation}
\end{subequations}
where the left-hand side of \eqref{eq:WHfinal} is analytic in the upper-half plane, and the right-hand side is analytic in the lower half plane. 
 
 After examining the asymptotic forms of both the left- and right-hand sides of \eqref{eq:WHfinal} above, i.e., after using the asymptotic forms \eqref{eq:asyjump}, \eqref{eq:asyfluidflow}, and \eqref{eq:Ksinfasy},  we find from Crighton et al.~\cite[pp 151]{crighton1992modern} that
\begin{subequations}
\begin{align}
\lim_{|s| \rightarrow \infty} A^+(s) &\sim \lim_{x\rightarrow 0} \int_0^\infty \phi^\rms(x,-d)\big|_{-}^{+}\, \rme^{\rmi s x}\, \rmd x \sim \frac{1}{s^{3/2}}, \\
\lim_{|s| \rightarrow \infty} B^-(s) &\sim \lim_{x\rightarrow 0} \int_{-\infty}^0  \partial_z \phi^\rms(x,-d) \, \rme^{\rmi s x} \, \rmd x \sim \frac{1}{s^{1/2}}, \\
\lim_{|s| \rightarrow \infty} K^\pm(s) &\sim s^{5/2},  
\end{align}
\end{subequations}
 and   using an extended form of Liouville's theorem (see Noble \cite[pp. 6]{noble1959methods}) yields  
\begin{subequations}
\label{eq:Whopfsyst}
 \begin{align}
\label{eq:WHalmost1}
 K^+(s) A^+(s) - \frac{\rmi}{(s+\ell)_+} \frac{1}{K^-(-\ell)}&= J(s), \\
\label{eq:WHalmost2}
 \frac{ P(s)}{K^-(s)} +(s^4 - \mu^4)   \frac{B^-(s)}{K^-(s)}  + \frac{\rmi}{(s+\ell)_+}\left(     \frac{1}{K^-(s)}   -  \frac{1}{K^-(-\ell)} \right)   &= J(s),
\end{align}
\end{subequations}
where $J(s) = ps + q$ and both $p$ and $q$ are unknown.   In order to construct the exact solution for the system above, we must determine the six unknown parameters that feature in the system above (i.e., the four unknowns from $P(s)$ and two unknowns from $J(s)$).

\section{Determining the polynomials P(s) and J(s)} \label{sec:dtpj}
Recall from   the half-range Fourier transform of the plate operator \eqref{eq:intbyparts4x}    that we obtain the entire function
\begin{equation}
P(s) =  (\rmi   \partial_z \phi^\rms_0 )s^3  - (\partial_x \partial_z \phi^\rms_0 )  s^2  - (\rmi  \partial_x^2 \partial_z \phi^\rms_0 ) s + (\partial_x^3 \partial_z \phi^\rms_0 ) 
\end{equation}
where subscript zero notation denotes the quantity as $x\rightarrow 0_-$. Any two of these constants are specified by the boundary conditions imposed at the plate edge, for example,    the most common boundary conditions take the form
\begin{subequations}
\label{eq:allbcsplate}
\begin{align}
\mbox{clamped edge $(w=0,\, \partial_x w = 0)$}: \qquad & \partial_z \phi^\rms_0 =  \frac{1}{\ell^4-\mu^4}, \quad \partial_x\partial_z \phi^\rms_0 =  \frac{\rmi \ell}{\ell^4-\mu^4}, \\
\mbox{simply supported edge $(w=0,\, \mathcal{M} = 0)$}: \qquad & \partial_z \phi^\rms_0 =  \frac{1}{\ell^4-\mu^4}, \quad \partial_x^2\partial_z \phi^\rms_0 = -\frac{\ell^2}{\ell^4-\mu^4}, \\
\mbox{free edge $(\mathcal{M}=0,\, \mathcal{V} = 0)$}: \qquad & \partial_x^2\partial_z \phi^\rms_0 = - \frac{\ell^2}{\ell^4-\mu^4}, \quad \partial_x^3\partial_z \phi^\rms_0 =  -\frac{\rmi \ell^3}{\ell^4-\mu^4},
\end{align}
\end{subequations}
where $\mathcal{M}=-D \partial_x^2 w$ is the bending moment  and  $\mathcal{V} = -D \partial_x^3 w$  the Kelvin--Kirchhoff edge reaction  of a thin elastic plate. Note that the absence of a variation in the $y$ direction leads to the  simplified forms above \cite[cf. pp. 4]{leissa1969vibration}.

\begin{figure}[t]
\centering
 
\includegraphics[width=0.70\textwidth]{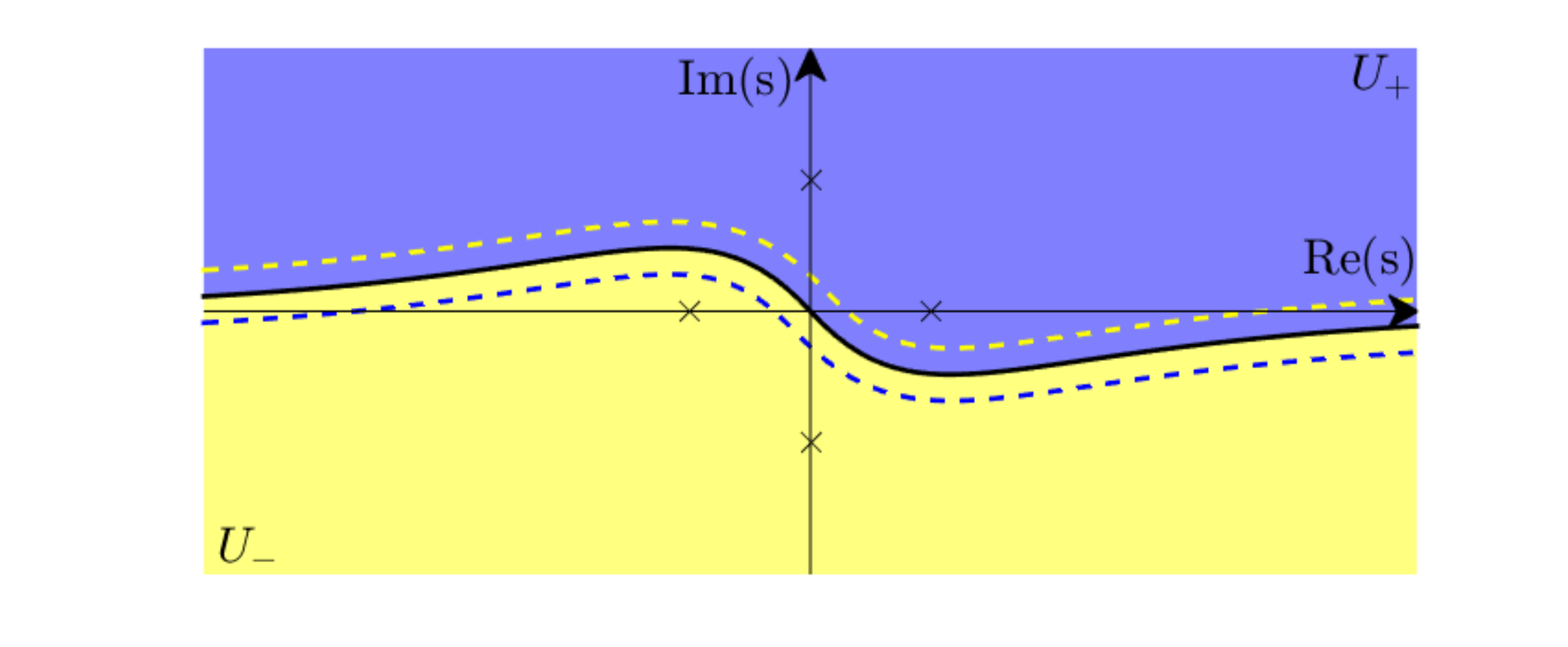}
 
\caption{\small Illustrative partition of complex plane into upper- and lower- half spaces ($U_+$ shaded in blue and $U_-$ in yellow respectively), for some function with simple poles at $s = \pm 1$ and $s = \pm\rmi $. Dashed lines outline  an  analytic continuation  of these two domains to form a  strip of analyticity enclosing the real line at $s = \pm \infty$. As we deform the partition (to have an interface along the real line) it indents below (above) any isolated singularities on the positive (negative) real line to ensure Sommerfeld radiation conditions are met when computing inverse Fourier transforms.
 \label{fig:all}}
  \end{figure}

However, after imposing the boundary condition at the edge, four unknown constants still remain. Following Cannell  \cite{cannell1975edge} we obtain two conditions   by rearranging the second equation in the partitioned Wiener--Hopf system \eqref{eq:WHalmost2} above to read
\begin{equation}
\label{eq:rearraWHalmost2}
  B^-(s)       = \frac{K^-(s)}{(s^4 - \mu^4)}\left[ ps + q -  \frac{\rmi}{(s+\ell)_+}\left(     \frac{1}{K^-(s)}   -  \frac{1}{K^-(-\ell)} \right) -  \frac{ P(s)}{K^-(s)}  \right].
\end{equation}
Since  $B^-(s)$ must be analytic in the lower-half plane, it follows from \eqref{eq:rearraWHalmost2} that 
\begin{equation}
\label{eq:condfornonsing1}
\left[ ps + q -  \frac{\rmi}{(s+\ell)_+}\left(     \frac{1}{K^-(s)}   -  \frac{1}{K^-(-\ell)} \right) -  \frac{ P(s)}{K^-(s)}  \right] = 0,
\end{equation}
when $s= -\mu$ and $s=-\rmi \mu$. For reference we designate the upper- and lower-half planes in Fig.~\ref{fig:all}.

Two further conditions are obtained by taking the half-range Fourier transform of the forced plate equation \eqref{eq:plate2}, to obtain 
\begin{equation}
M^+(s) =  \frac{1}{(s^4 - \mu^4)}\left[ P(s) + \frac{\rmi}{(s+\ell)_+} - \beta A^+(s)\right],
\end{equation}
after using the decomposition for $\partial_z \phi^\rms$ in \eqref{eq:phizbx}.
Since $M^+(s)$ must be analytic in the upper-half plane, it follows that
\begin{equation}
\label{eq:condfornonsing2}
\left[ P(s) + \frac{\rmi}{(s+\ell)_+} - \beta A^+(s)\right] = \left[ P(s) + \frac{\rmi}{(s+\ell)_+} -  \frac{\beta}{K^+(s) }\left(    ps + q + \frac{\rmi}{(s+\ell)_+} \frac{1}{K^-(-\ell)} \right)\right]= 0,
\end{equation}
when $s= \mu$ and $s=\rmi \mu$. Closed-form expressions for the unknown constants are not given here as they are  quite lengthy. Thus, the Wiener--Hopf system has now been solved to give the functions $A^+(s)$, $B^-(s)$, and $M^+(s)$ explicitly.

\section{Solution representation} \label{sec:solrep}
 Having determined the necessary polynomials above, we return to \eqref{eq:WHalmost1} to write
 \begin{equation}
A^+(s)=   \frac{1}{K^+(s) }\left(    ps + q + \frac{\rmi}{(s+\ell)_+} \frac{1}{K^-(-\ell)} \right).
  \end{equation}
Using the relationship between $H^\rmL(s)$ and $A^+(s)$ in \eqref{eq:fouriertransApD}, the relationship between $G^\rmU(s)$ and $H^\rmL(s)$ in \eqref{eq:CsvsDs}, the general solution for $\Psi(s,z)$ in \eqref{eq:psipartition} is now completely prescribed. Hence, we write the total field in $x<0$ as  
\begin{subequations}
\begin{equation} 
\mathcal{A}\phi(x,z) =\phi^\mathrm{inc}(x,z)+
  \frac{1}{2\pi}\oint_{U_+} \Psi(s,z) \, \rme^{-\rmi s x}\, \rmd s, 
\label{eq:phireflasdf}
\end{equation}
where  the integral indicates a closure in the upper-half plane, which is evaluated as
\begin{equation}
\mathcal{A}\phi(x,z)  =\phi^\mathrm{inc}(x,z)+  
   \sum_j \rme^{-\rmi s_j^\mathcal{R} x} 
\begin{cases} 
       \tau^\rmU(s_j^\mathcal{R})       \left[  \cosh(s_j^\mathcal{R} z) +\dfrac{\alpha}{s_j^\mathcal{R}} \sinh(s_j^\mathcal{R} z) \right]  &  \mbox{for }  -d<z \leq 0, \\
\\
   \tau^\rmL(s_j^\mathcal{R})      \cosh(s_j^\mathcal{R}(z+h))      & \mbox{for }  -h\leq z<-d, 
   \end{cases}  
\end{equation}
in which
\begin{align}
\tau^\rmU(s) &=  \frac{\rmi}{K^+(s) }\left(    ps + q + \frac{\rmi}{(s+\ell)_+} \frac{1}{K^-(-\ell)} \right)  \frac{ s \sinh(s c) }{ (1-\alpha s) \sinh(s h) +  s^2 \cosh(s h)} , \\
\tau^\rmL(s) &= \frac{\rmi}{K^+(s) }\left(    ps + q + \frac{\rmi}{(s+\ell)_+} \frac{1}{K^-(-\ell)} \right)  \frac{ \eta(s)}{ (1-\alpha s ) \sinh(s h) + s^2 \cosh(s h)} ,
\end{align}
\end{subequations}
and   $s_j^\mathcal{R}$ denote solutions to $  s  \sinh(sh) -\alpha \cosh(sh) =0$ in $U_+$, which are all simple poles.  Since there is no discontinuity for $x<0$ the above expression can be simplified to 
\begin{equation}
\mathcal{A}\phi(x,z)  =\phi^\mathrm{inc}(x,z)+  
   \sum_j \rme^{-\rmi s_j^\mathcal{R} x} \,
   \tau^\rmL(s_j^\mathcal{R})      \cosh(s_j^\mathcal{R}(z+h))    \,\,  \mbox{for }  -h\leq z<0.  
\end{equation}

Likewise   the total field in $x>0$ takes the form  
\begin{equation}
\label{eq:phitrans}
\mathcal{A}\phi(x,z) = 
 \phi^\mathrm{inc}(x,z)+
 \frac{1}{2\pi}\oint_{U_-}  \Psi(s,z) \, \rme^{-\rmi s x}\,  \rmd s ,
\end{equation}
where the integral is closed in the lower-half plane, which is evaluated via
\begin{align}
\mathcal{A}\phi(x,z) =  \sum_j   \rme^{\rmi  s_j^\mathcal{T} x}   
 \begin{cases} 
          \kappa^\rmU( -s_j^\mathcal{T}) \left[  \cosh( s_j^\mathcal{T} z) +\dfrac{\alpha}{ s_j^\mathcal{T}} \sinh( s_j^\mathcal{T} z) \right]  &  \mbox{for }  -d<z \leq 0, \\
\\
      \kappa^\rmL( -s_j^\mathcal{T}) \cosh( s_j^\mathcal{T}(z+h))      & \mbox{for }  -h\leq z<-d,
   \end{cases}    
\end{align}
in which
\begin{align}
\kappa^\rmU(s) &=   -\rmi  K^-(s) \left(    (ps + q)(s+\ell)_+  + \frac{\rmi}{K^-(-\ell)} \right) \frac{s \sinh(s c) }{\partial_s y(s)}      , \\
\kappa^\rmL(s) &=  -\rmi  K^-(s) \left(    (ps + q)(s+\ell)_+  + \frac{\rmi}{K^-(-\ell)} \right)  \frac{\eta(s)}{\partial_s y(s)} ,
\end{align}
and
\begin{equation}
y(s) =\left[s(s^4-\mu^4) \sinh(sc) \eta(s) + \beta\left\{s \sinh(sh)-\alpha \cosh(sh)\right\} \right] (s+\ell)_+.
\end{equation}
In the above, $ s=-s_j^\mathcal{T}$ are solutions to $y(s) = 0$ in $U_-$, excluding $s_j^\mathcal{T} = \ell$, and   are all simple poles.
 Note that the sign of the residue is due to the clockwise orientation taken in the lower-half plane. Furthermore, both sums are evaluated over both real, strictly imaginary, and complex-valued roots, to ensure that the evanescent field contributions are included, and the entire field is appropriately reconstructed close to the plate edge.
 
\subsection{Conservation of energy relation}
As outlined in existing works \cite{crighton1991fluid,ulhassan2009water,williams2012wiener} we use Green's second identity with the fluid potential and its conjugate to obtain the energy-balance expression
\begin{equation}
\label{eq:ebalplate}
|\mathcal{A}|^2\mathrm{Im} \left\{ \oint_{\partial \mathcal{V}} \phi^\ast \partial_n \phi \, dS  \right\}= 0,
\end{equation}
where $\mathcal{V} = \left\{  \left\{-x_\rmW<x<x_\rmW  \right\}\times \left\{ -h<z<0 \right\} \right\}\backslash \Gamma_\rmp$. 
For $x_\rmW$ large, it follows that only propagating waves are present, i.e., the incident and reflected surface wave on the left 
\begin{subequations}
\begin{align}
\lim_{x\rightarrow -\infty}\mathcal{A}\phi(x,z) \sim  \phi^\mathrm{inc}(x,z) +  \tau^\rmL(\ell) \cosh(\ell(z+h)) \,  \rme^{-\rmi  \ell x},    & \quad \mbox{for }  -h\leq z<0,  
\end{align}
and both the transmitted surface wave and coupled plate-surface wave on the right
\begin{align}
\lim_{x\rightarrow  \infty} \mathcal{A}\phi(x,z) \sim
 \begin{cases} 
        \kappa^\rmU(-k_1) \left[  \cosh(k_1 z) +\dfrac{\alpha}{k_1} \sinh(k_1 z) \right] \rme^{\rmi k_1 x} \\   \hspace{15mm}+
\kappa^\rmU(-k_2) \left[  \cosh(k_2 z) +\dfrac{\alpha}{k_2} \sinh(k_2 z) \right] \rme^{\rmi k_2 x} &  \mbox{for }  -d<z \leq0, \\
\\
   \kappa^\rmL(-k_1) \cosh(k_1(z+h)) \,  \rme^{\rmi k_1 x} 
   \\   \hspace{15mm}+
  \kappa^\rmL(-k_2) \cosh(k_2(z+h)) \,  \rme^{\rmi k_2 x}    & \mbox{for }  -h\leq z<-d,  
   \end{cases}  
\end{align}
\end{subequations}
where $\tau^{\rmU,\rmL}$ and $\kappa^{\rmU,\rmL}$ are residue contributions to the amplitudes, and $k_j = s_j^\mathcal{T}$ for convenience. After imposing the plate boundary conditions at the edge we obtain the energy-balance relation
\begin{multline}
\label{eq:enbalancezerop}
-\ell \int_{-h}^{0} \frac{\cosh^2(\ell(z+h))}{\ell^2(\ell^4-\mu^4)^2 \sinh^2(\ell c)} \, \rmd z  
+ \ell |\tau^\rmL_\ell|^2 \int_{-h}^{0}  \left[ \cosh(\ell(z+h)) \right]^2 \,\rmd z
\\
+  k_1 |\kappa_1^\rmU|^2\int_{-d}^{0}  \left[  \cosh(k_1 z) +\dfrac{\alpha}{k_1} \sinh(k_1 z) \right] ^2 \,\rmd z \\
+ k_2 |\kappa_2^\rmU|^2 \int_{-d}^{0}  \left[  \cosh(k_2 z) +\dfrac{\alpha}{k_2} \sinh(k_2 z) \right] ^2 \,\rmd z
+ k_1 |\kappa_1^\rmL|^2 \int_{-h}^{-d}   \cosh^2(k_1(z+h)) \, \rmd z \\
+k_2 |\kappa_2^\rmL|^2  \int_{-h}^{-d}   \cosh^2(k_2(z+h)) \, \rmd z 
- \frac{2}{\beta} \left[ | \kappa_1^\rmL|^2 k_1^5 \sinh^2(k_1 c) + | \kappa_2^\rmL|^2 k_2^5 \sinh^2(k_2 c)\right] = 0,
\end{multline}
where $\kappa_{1,2}^\rmL=\kappa^\rmL(-k_{1,2})$ refer to the two waves propagating towards $x\rightarrow \infty$ and the final term arises from integration over the plate, i.e., 
\begin{multline}
\label{eq:elplatecontribut}
\mathrm{Im} \left\{ \int_{0}^{x_\rmW} (\phi(x,-d)\big|^+_-)^\ast \partial_z \phi(x,-d) \, \rmd x  \right\}= 
-\frac{1}{\beta} \mathrm{Im} \left\{ \int_{0}^{x_\rmW} (\partial_x^4 \partial_z \phi(x,-d) )^\ast \partial_z \phi(x,-d) \, \rmd x   \right\} \\
=-\frac{1}{\beta} \mathrm{Im} \left\{ \left[   \left( \partial_x^3 \partial_z \phi(x,-d) \right)^\ast \partial_z \phi(x,-d)  -
\left(\partial_x^2 \partial_z \phi(x,-d) \right)^\ast \partial_x \partial_z \phi(x,-d)  \right]_0^{x_\rmW} \right\} \\
= -\frac{2}{\beta} \left[| \kappa_1^\rmL|^2 k_1^5 \sinh^2(k_1 c) + |\kappa_2^\rmL|^2 k_2^5 \sinh^2(k_2 c)\right].
\end{multline}
The energy-balance equation \eqref{eq:enbalancezerop} may be expressed in a more familiar  form through the scaling 
\begin{equation}
\mathcal{A} = \frac {-\cosh(\ell h))}{\ell (\ell^4 - \mu^4)  \sinh(\ell c) },
\label{A_equation}
\end{equation}
from which the asymptotic form of the potential is  
\begin{equation}
 \phi(x,0)   \sim
 \begin{cases}
        \rme^{\rmi \ell x}+ R\rme^{-\rmi \ell x}, & x\rightarrow -\infty \\
        T_1\rme^{\rmi k_1 x} + T_2 \rme^{\rmi k_2 x}, & x\rightarrow \infty
 \end{cases}
 \label{far_field}
\end{equation}
where the reflection coefficient is given by
\[
R = -\tau^\rmL(\ell) \ell (\ell^4 - \mu^4)  \sinh(\ell c),
\]
and 
\[
T_j = -\frac{\ell (\ell^4 - \mu^4)  \sinh(\ell c) }{\cosh(\ell h))}\kappa^\rmU(-k_j).
\]
It is then clear that the energy-balance equation can be written in terms of reflection and transmission coefficients and it is equivalent to that given in ul-Hassan et al.~\cite[Eq.~(C.10)]{ulhassan2009water}. 

\subsection{Numerical Examples}
  In this section, we compute the response of a submerged elastic plate configuration following the procedure outlined in the preceding sections. We also present the total reflection and total transmission coefficients over a wide frequency range as a benchmark for numerical validation in future studies. The numerical solution was validated by the energy-balance relation and by comparison with the previous solution via eigenfunction matching \cite{ulhassan2009water}.
  
\begin{figure}[t]
       \includegraphics[width=0.45\textwidth]{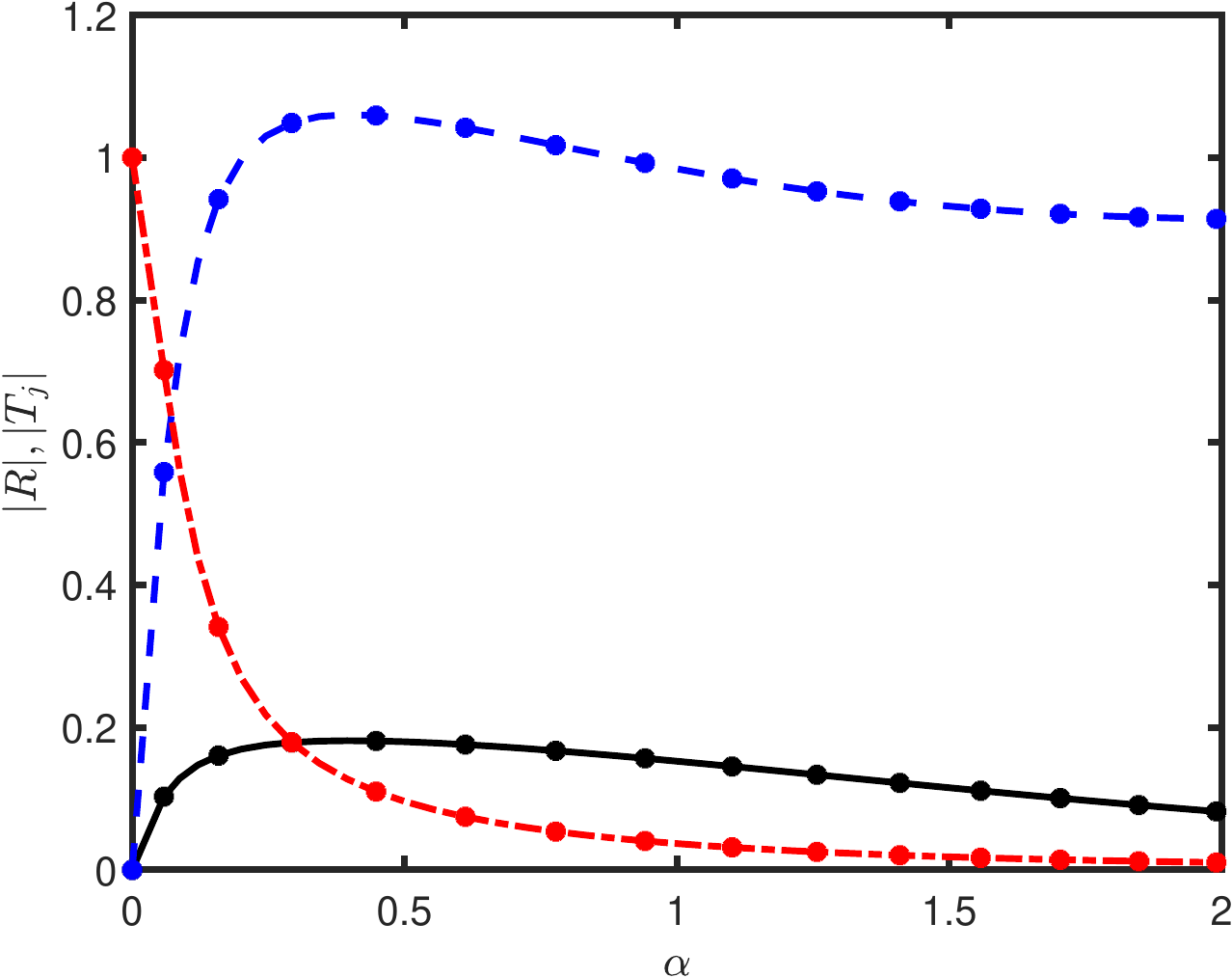}
     \hfill
       \includegraphics[width=0.45\textwidth]{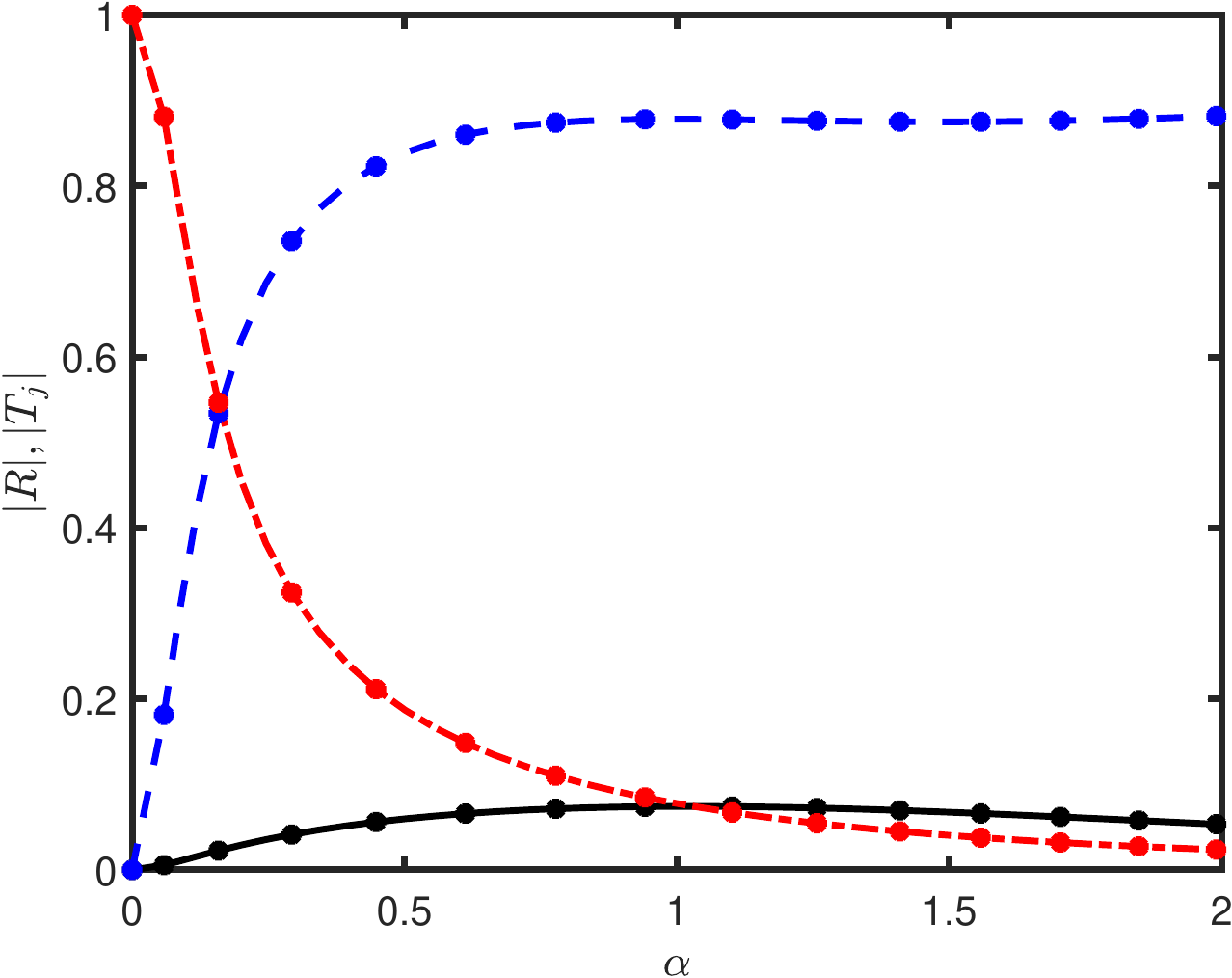}
     \caption{\small The reflection coefficient (black solid line), high wavenumber transmitted mode (red dot-dashed line) and
       low wavenumber transmitted mode (blue dashed line) for a submerged elastic plate where  $\mu = \alpha\times 10^{-3}$, $\beta = \alpha\times 10^{-2}$,  $h=1.5$ and $d=0.5$.
       Figure (a) corresponds to clamped edge and (b) free edge conditions at the plate edge. We superpose circles corresponding to results obtained using the eigenfunction matching method  \cite{ulhassan2009water}.
       }\label{fig:randt}
     \end{figure}

 In this section we rescale the incident field \eqref{eq:pwansantz2} so that its potential amplitude is unity at the free surface. Therefore we set $\mathcal{A}$ according to equation~\eqref{A_equation}. The scattered far field surface potential consists of a reflected wave of the form $R\mathrm{e}^{-\mathrm{i} \ell x}$ for large negative $x$  and two transmitted waves (as in equation~\eqref{far_field}). The two transmitted waves have different wavenumbers and we denote them by the high wavenumber and low wavenumber mode. The surface potential in the far field for large positive $x$ is $T_1\mathrm{e}^{ \mathrm{i} k_1 x} + T_2\mathrm{e}^{ \mathrm{i} k_2 x}$. Note here that $\ell$, $ k_1$, and $ k_2$ are the only purely real solutions. 
 These two far-field transmitted waves are analogous to the two waves found for a stratified fluid (the surface and internal or interface wave).  
 In Figure~\ref{fig:randt} we choose $h=1.5$ and $d=0.5$ which correspond to the values chosen by ul-Hassan et al.~\cite{ulhassan2009water}. We express the solution in terms of $\alpha = \omega^2/g$ and set $\mu = \alpha\times 10^{-3}$ and $\beta = \alpha\times 10^{-2}$.
 Figure~\ref{fig:randt} shows the absolute value of the reflection and transmission coefficients as a function of $\alpha$. We consider the two edge conditions, clamped and free.  The free edge condition problem has a significantly lower reflection over this frequency range. We also see that for both low and high frequencies, the transmission becomes unity. For reference, we also superpose results obtained using eigenfunction matching (as circular points), which match results for the reflection coefficient presented in Figures 3 and 4 of ul-Hassan et al.~\cite{ulhassan2009water}.

     For the remaining figures, we choose $h=2$ and $d=0.5$, $\mu = \alpha\times 10^{-2}$ and $\beta = \alpha\times 10^{-1}$. 
     We extend the monochromatic plane--wave solution to consider a time-dependent Gaussian wave pulse. We consider the surface displacement ($\eta$),  which we write as
     \[ \eta(x,t) = \mathrm{Re} \left\{ \int_0^{\infty} \hat{f}(k) \left.\partial_z\Phi(x,z;\omega(k))\right|_{z=0} \mathrm{e}^{-\mathrm{i} \omega(k) } \mathrm{d} k
     \right\},
     \]
     where $\hat{f}(k) = 5/(6\sqrt{\pi})\exp((k-k_0)^2/25)$ is a Gaussian centred at 
     $k_0 = 0.5$ where the $5/(6\sqrt{\pi})$ factor is arbitrary, but chosen so that the maximum pulse displacement, if the submerged plate was absent, is $1/6$. We found this value to give the best illustration of the fluid and plate motion.
     We choose the scaling factor $\mathcal{A}$ so that the normal derivative has unit amplitude at the free surface  
     \begin{equation*}
\mathcal{A} = \frac {-\sinh(\ell h)}{(\ell^4 - \mu^4)\sinh(\ell c)}. 
\end{equation*}
Similarly,  the displacement of the submerged plate is given by
 \[ w_b(x,t)= \mathrm{Re} \left\{ \int_0^{\infty} \hat{f}(k) \left.\partial_z\Phi(x,z;\omega(k))\right|_{z=-d} \mathrm{e}^{-\mathrm{i} \omega(k) } \mathrm{d} k
     \right\},\,\,x>0.
     \]
     The solution then appears as a reflected and two transmitted pulse each of which is travelling at a different group speed as can be seen in Figure~\ref{time_dependent_clamped} and the movie which can be found in the Supplementary Material. Note that we superpose the displacement of the plate relative to the rest position at $z=-d$ for illustrative purposes.
     \begin{figure}
\begin{center}
\includegraphics[scale = 0.4]{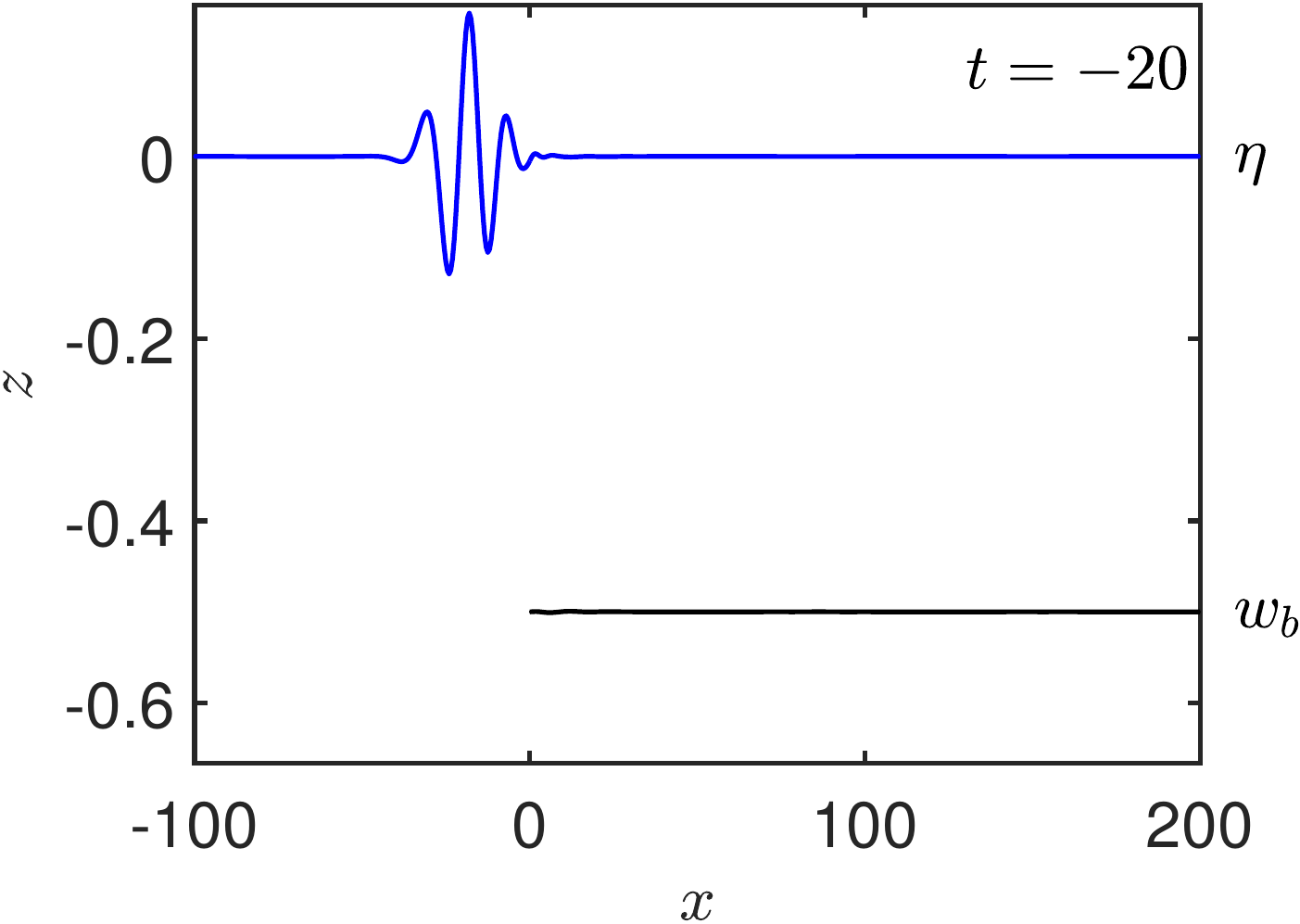}
\hspace{0.5cm}
\includegraphics[scale = 0.4]{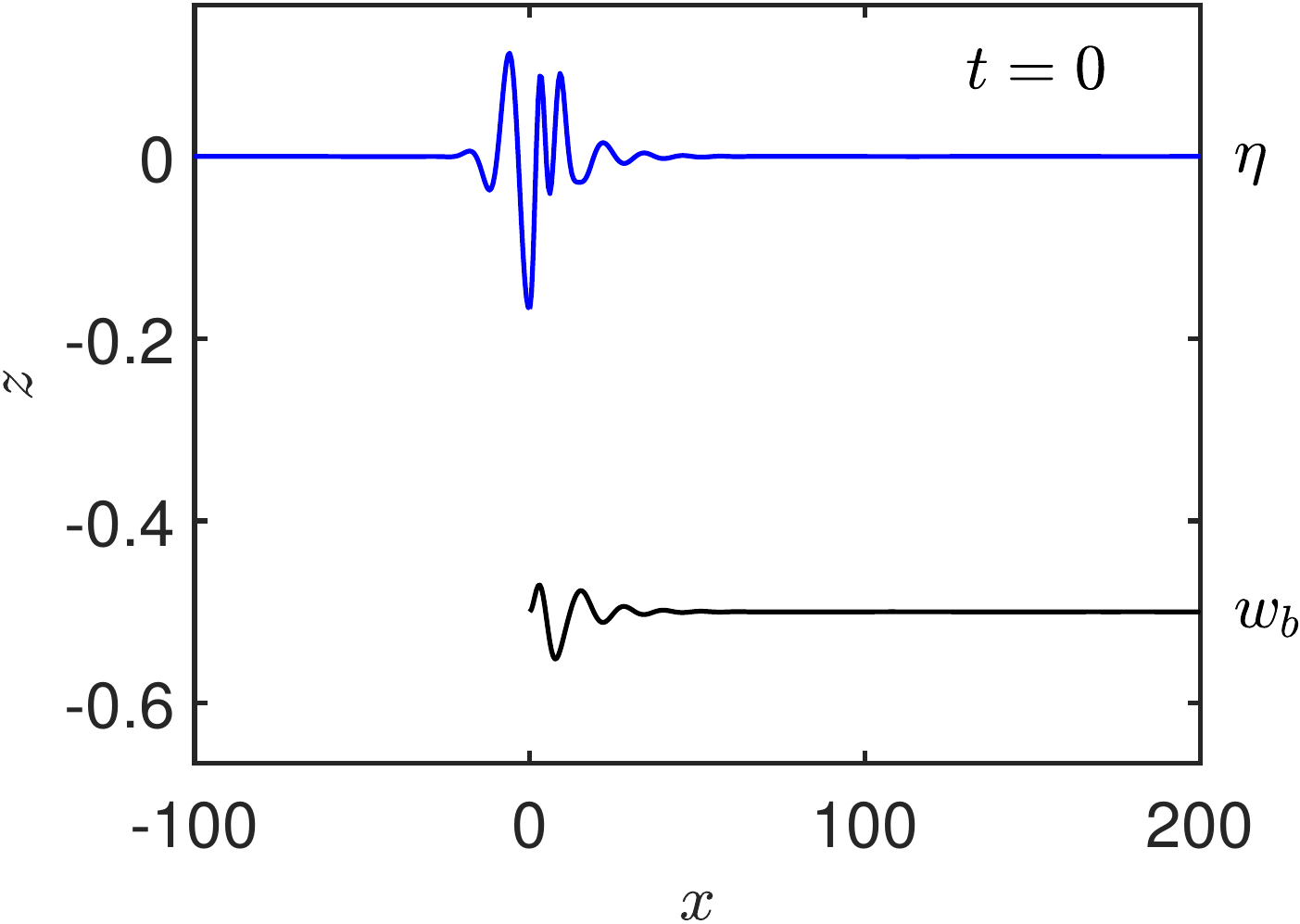}\\
\includegraphics[scale = 0.4]{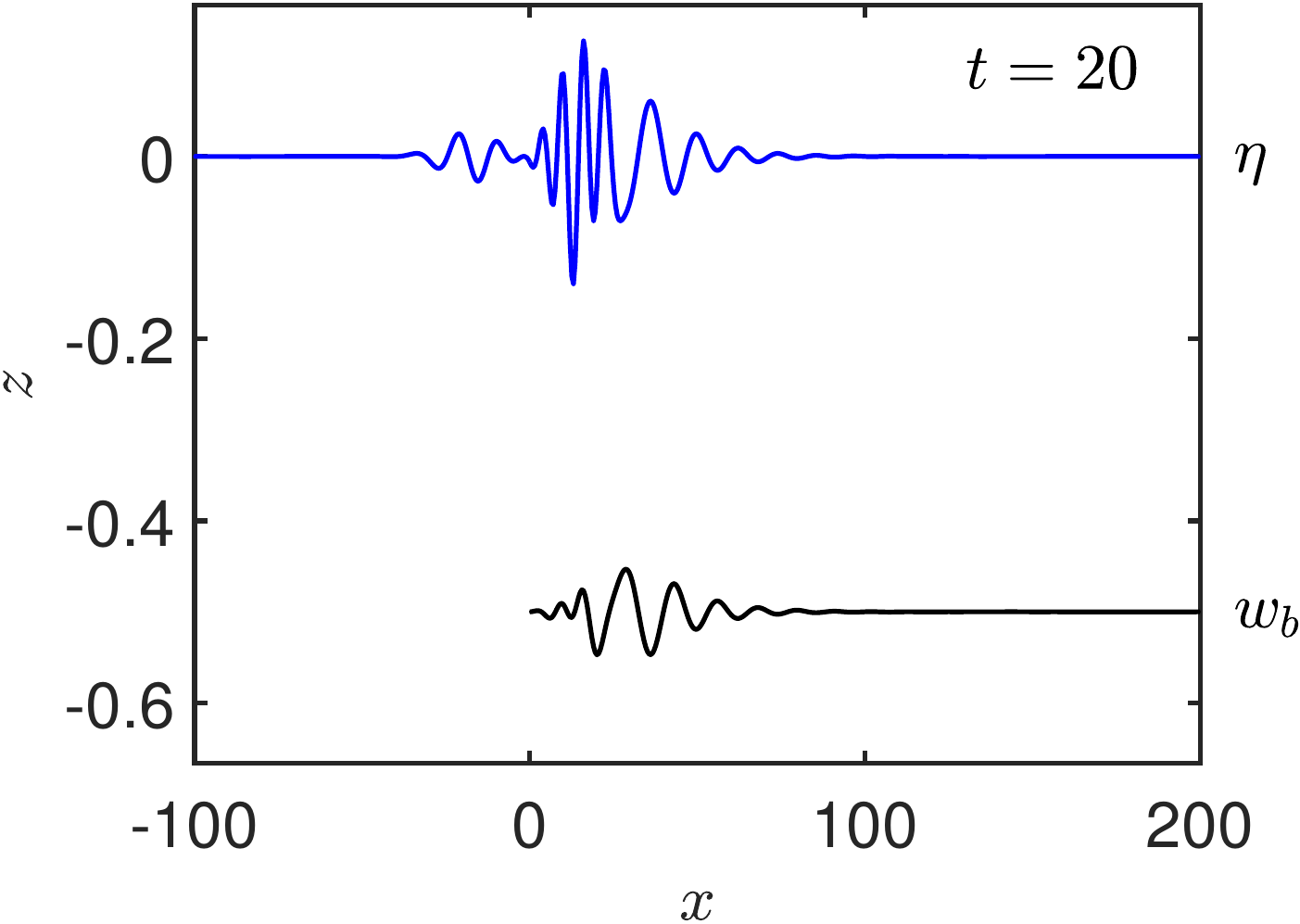}
\hspace{0.5cm}
\includegraphics[scale = 0.4]{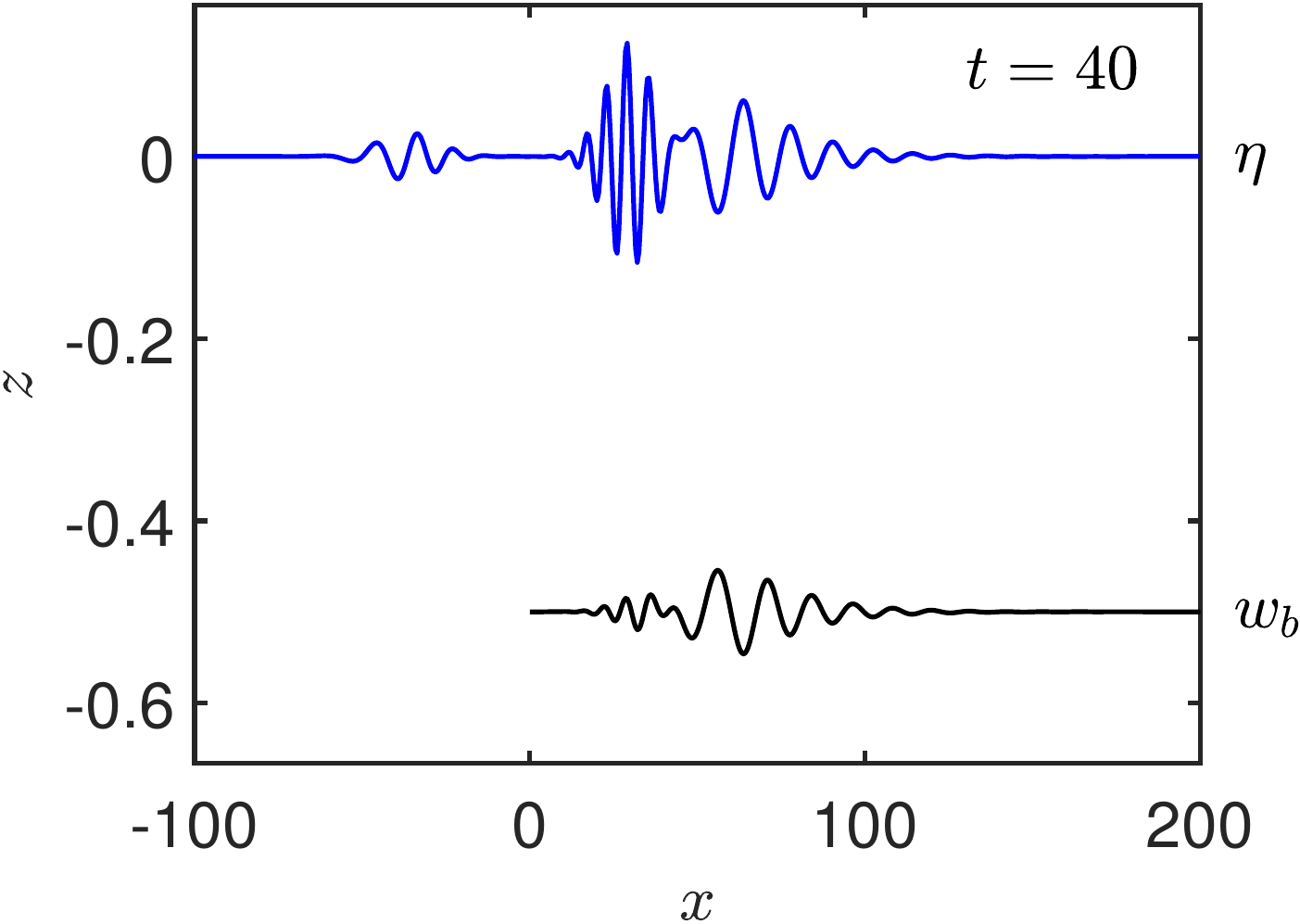}\\
\includegraphics[scale = 0.4]{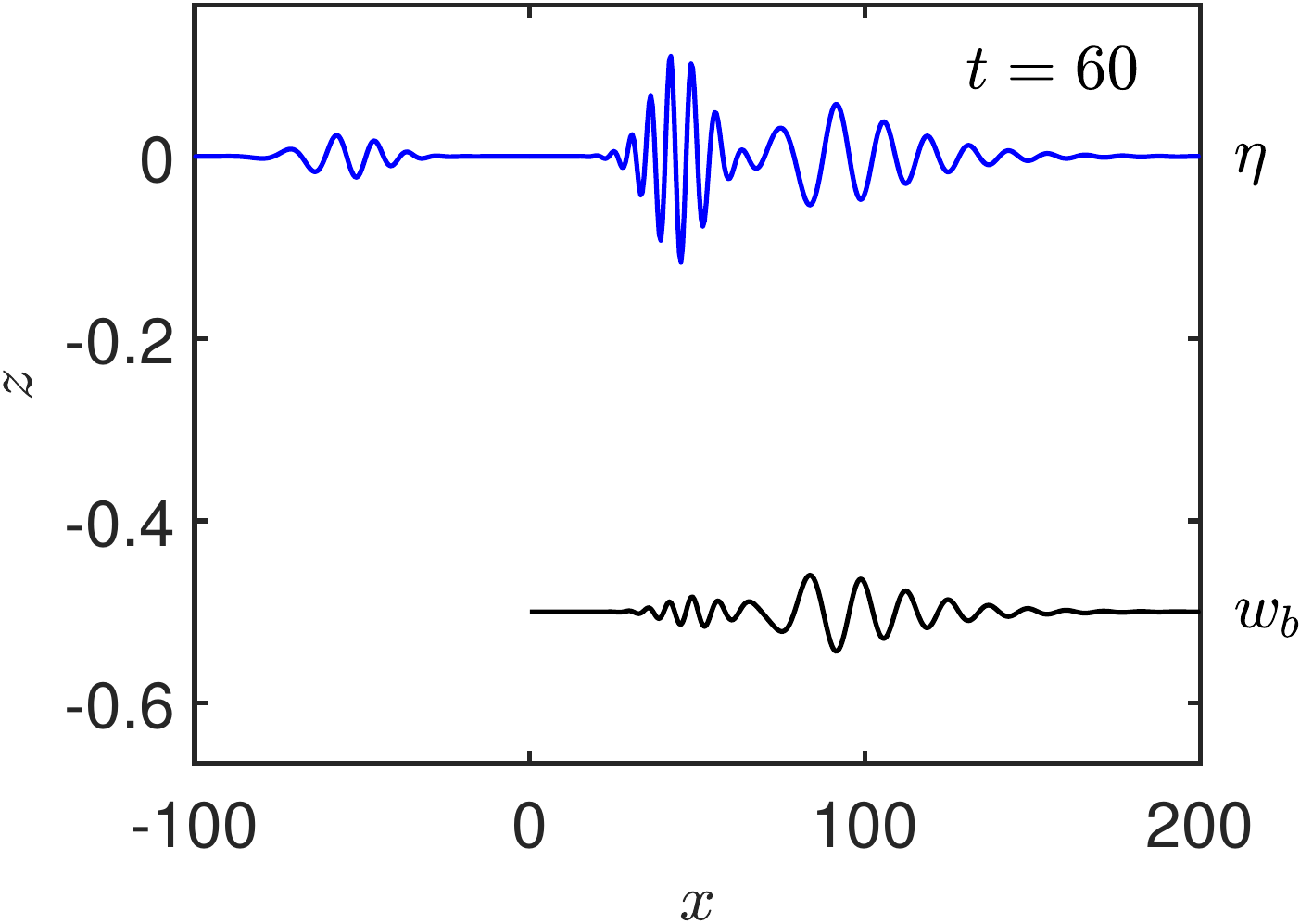}
\hspace{0.5cm}
\includegraphics[scale = 0.4]{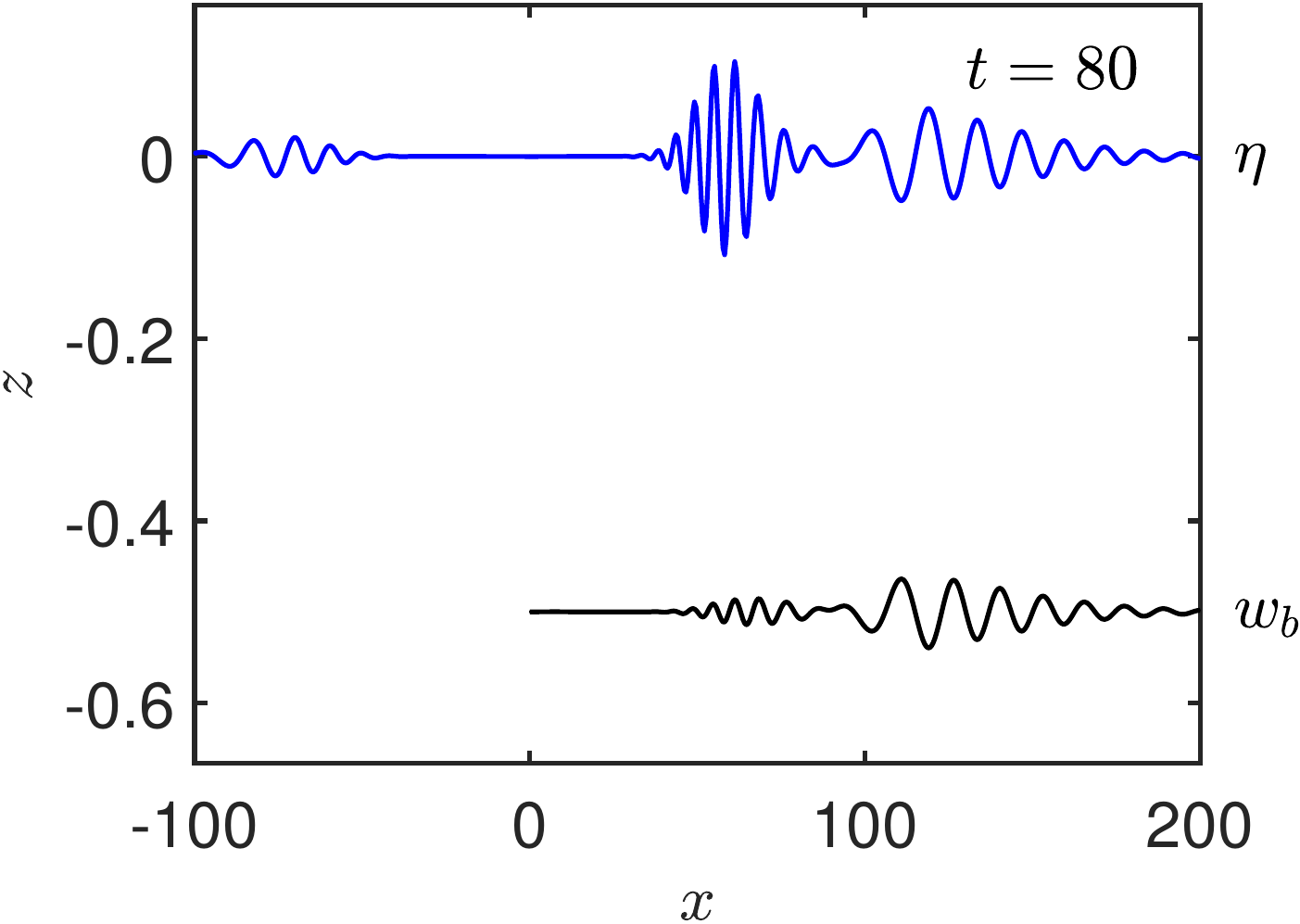}
\end{center}
\caption{\small The time-dependent motion of the fluid surface ($\eta$) in blue and submerged elastic plate ($w_b$), relative to its rest position, in black with clamped leading edge for a Gaussian incident wave pulse for parameter values  $\mu = \alpha\times 10^{-2}$ and $\beta = \alpha\times 10^{-1}$.    The full animation can be
found in movie 1 in the Supplementary Material. }%
\label{time_dependent_clamped}
\end{figure}

     \begin{figure}
\begin{center}
\includegraphics[scale = 0.4]{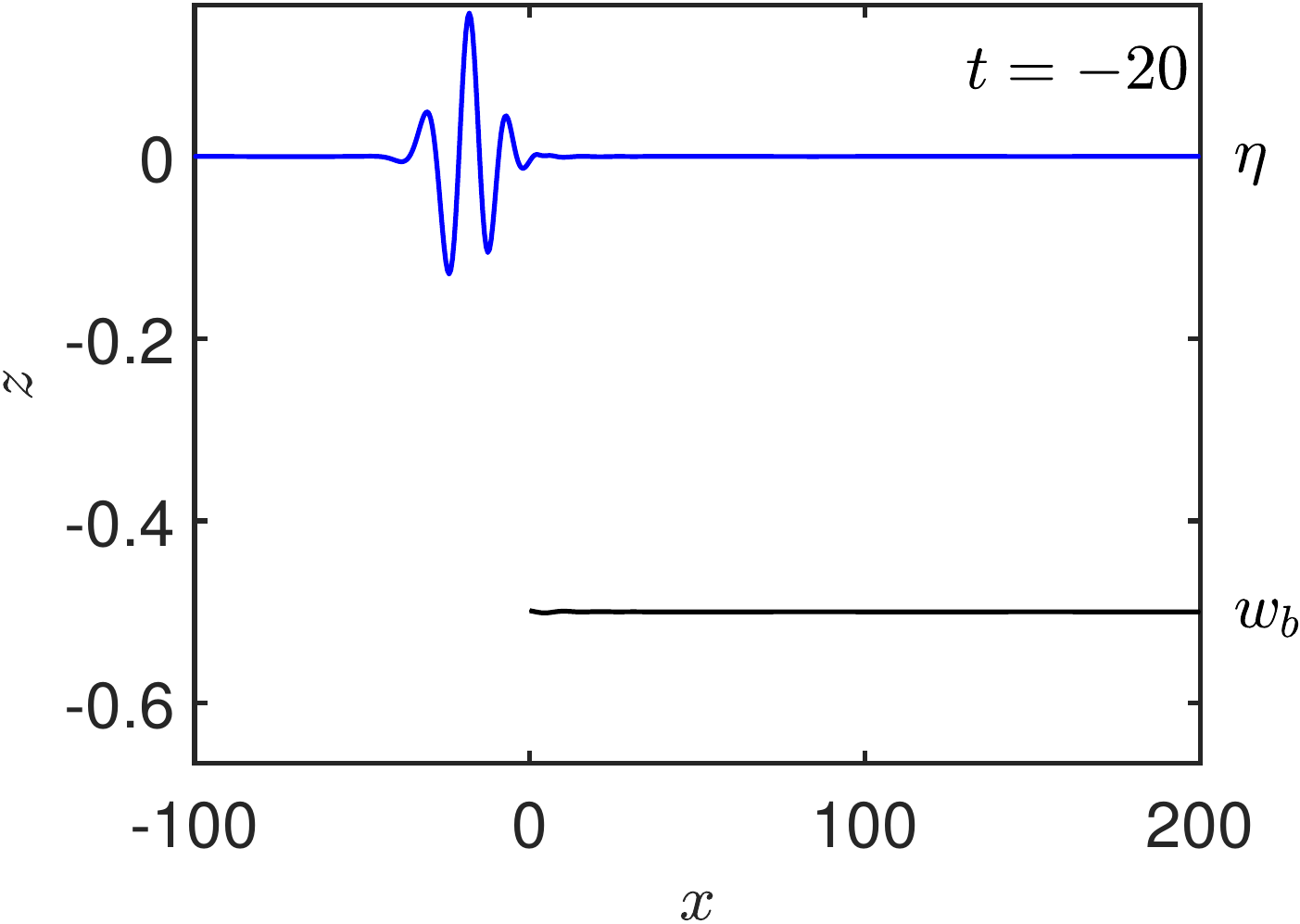}
\hspace{0.5cm}
\includegraphics[scale = 0.4]{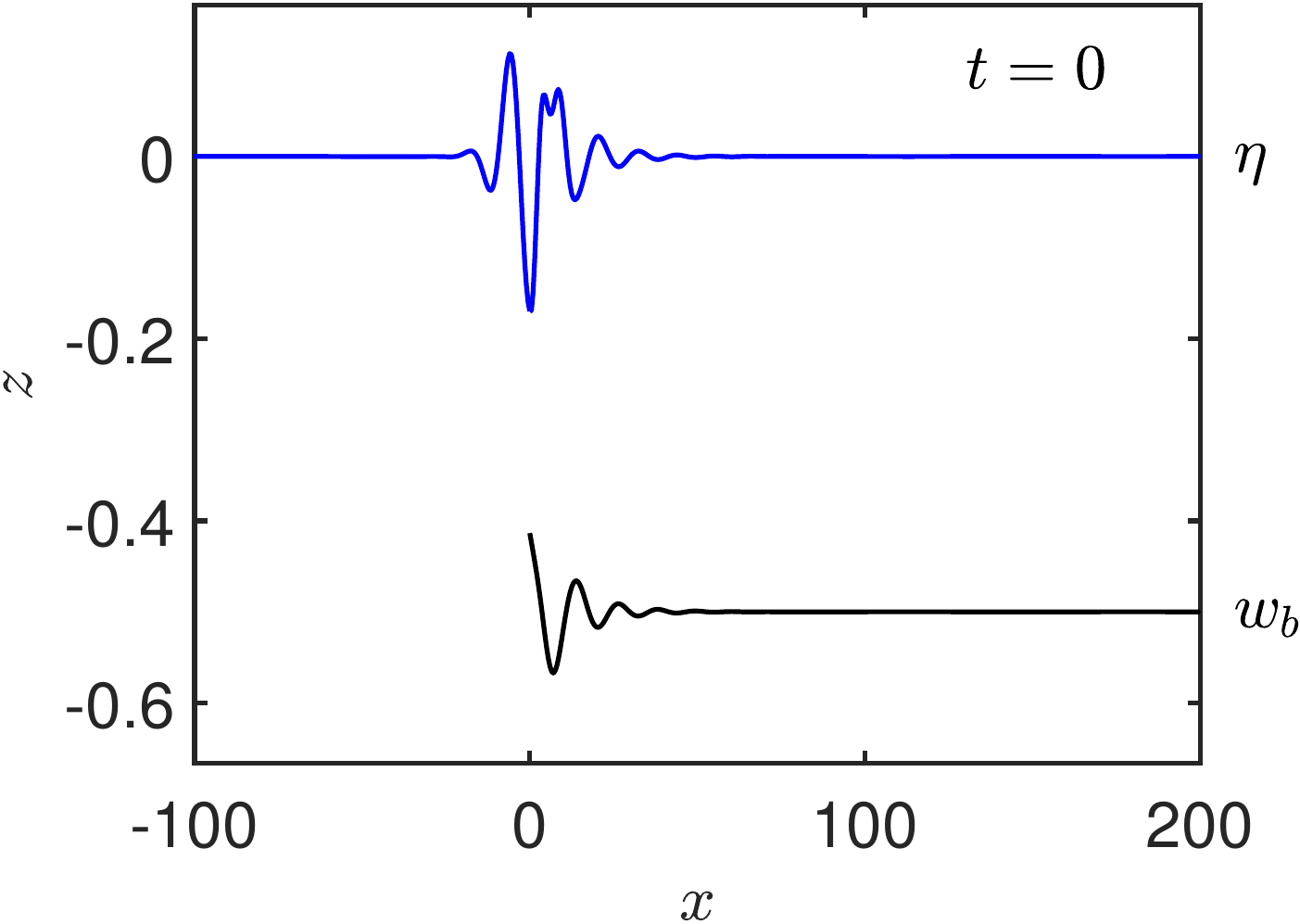}\\
\includegraphics[scale = 0.4]{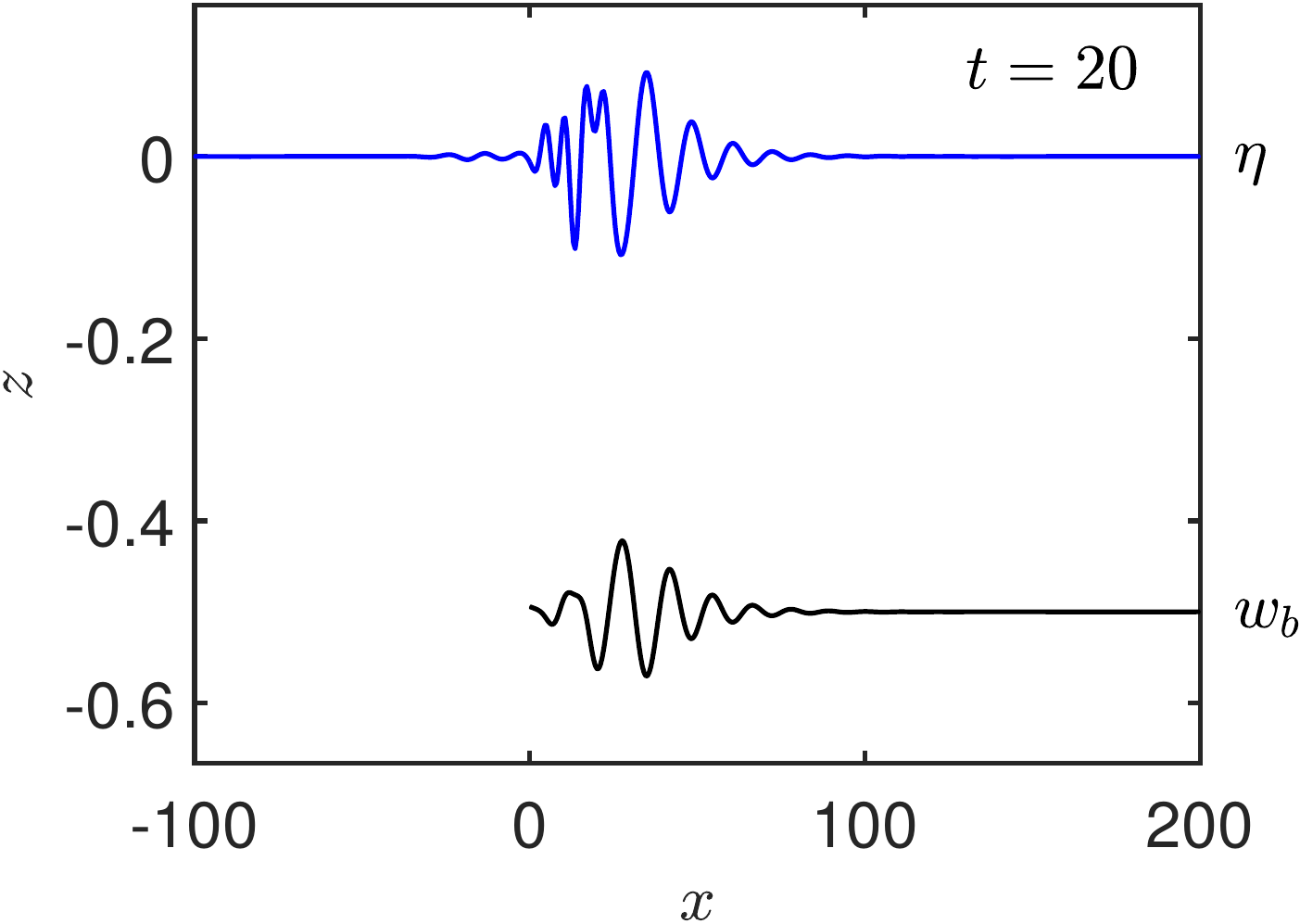}
\hspace{0.5cm}
\includegraphics[scale = 0.4]{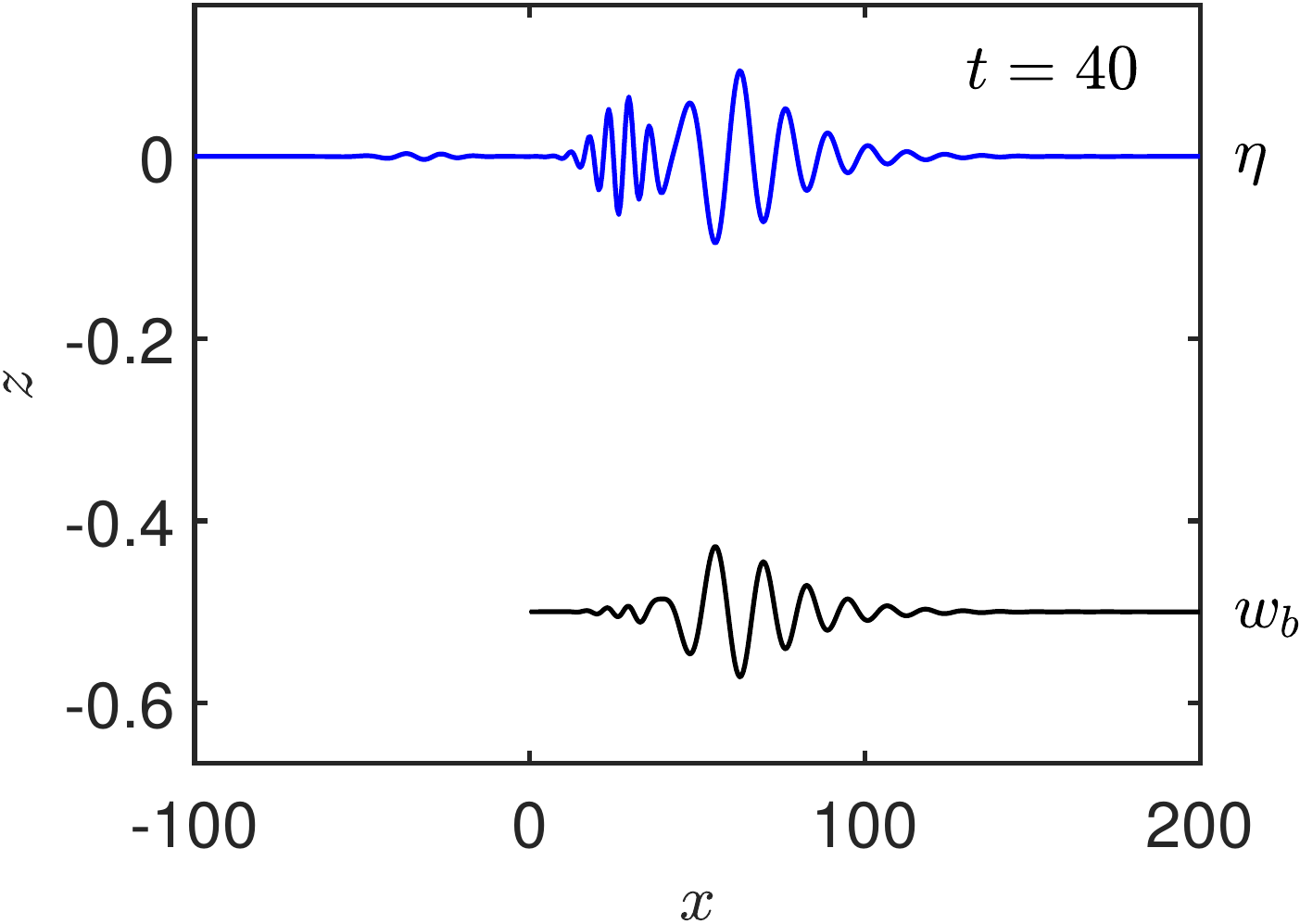}\\
\includegraphics[scale = 0.4]{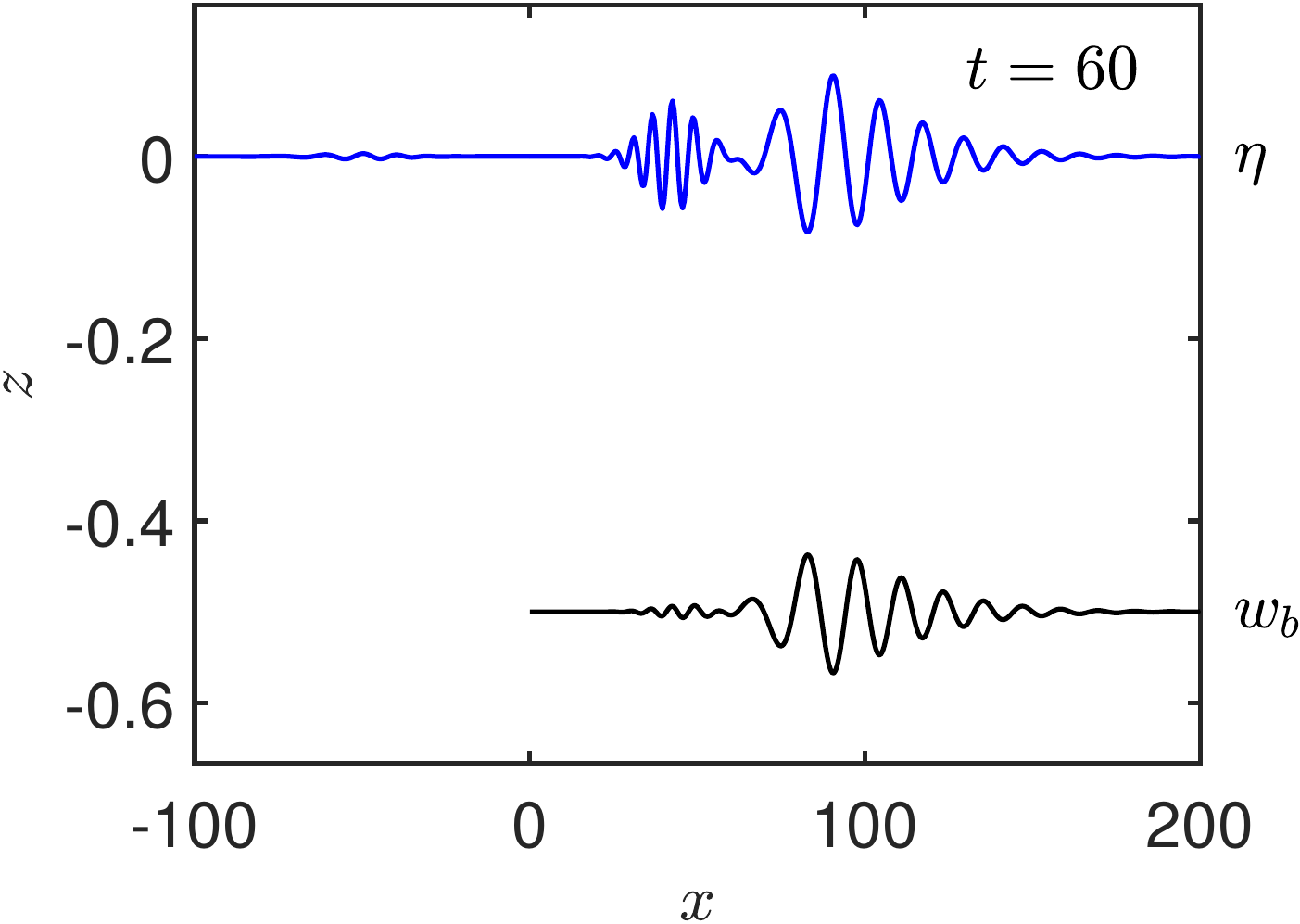}
\hspace{0.5cm}
\includegraphics[scale = 0.4]{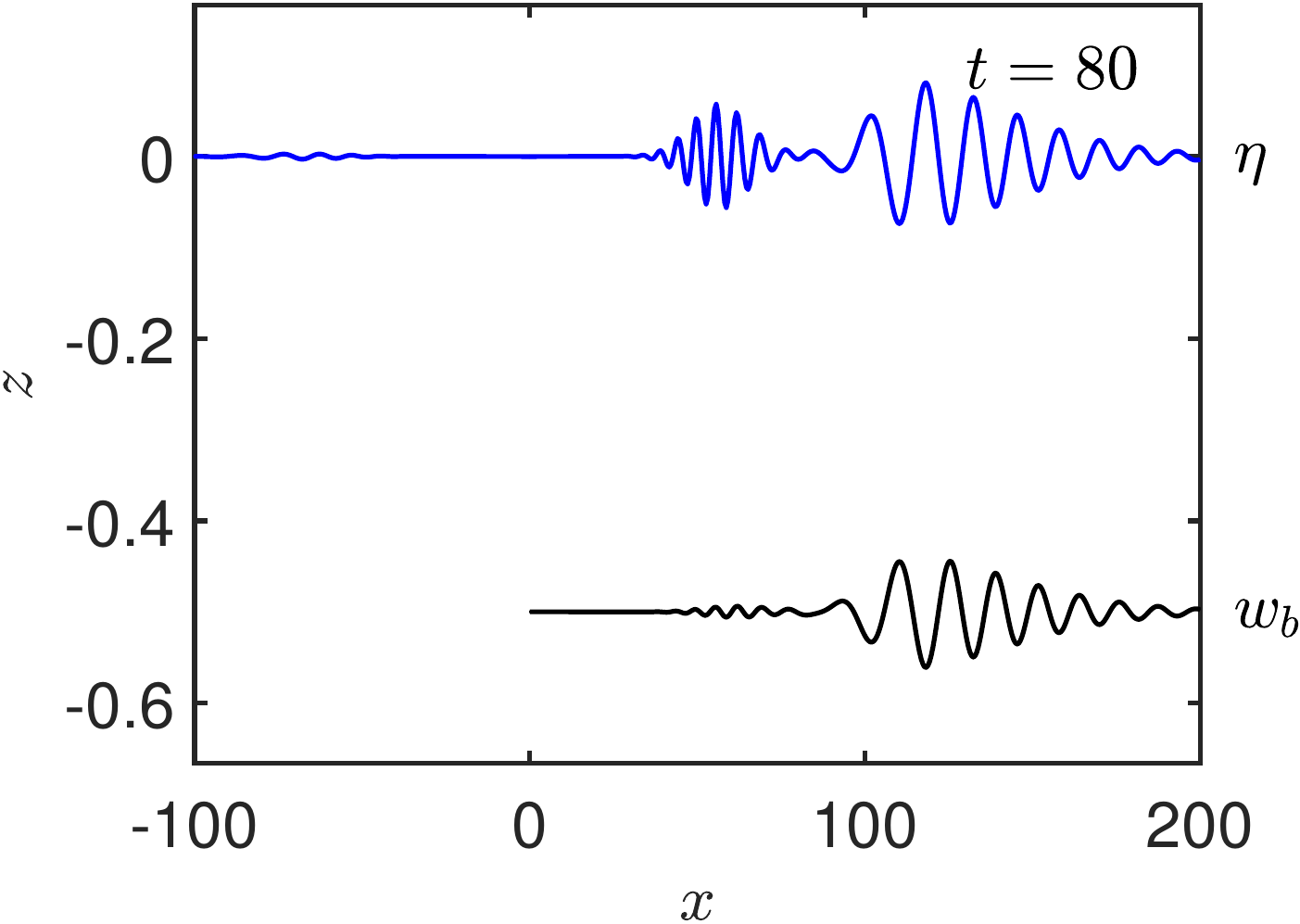}
\end{center}
\caption{\small As in Figure~\ref{time_dependent_clamped} except   with free edge conditions   applied at the plate edge. The full animation can be found in movie 2 in the Supplementary Material.}%
\label{time_dependent_free}
\end{figure}
 
\section{Extension to submerged poroelastic plates} \label{sec:extporo}
Next, we construct a poroelastic thin plate by drilling many small holes into an otherwise uniform thin elastic plate, for example, a Kirchhoff--Love plate described by \eqref{eq:kirchofflove}. Porous plates are widely used in subsea applications as they provide two mechanisms for impacting wave energy in a fluid medium: through the excitation of flexural modes and dissipative processes when transporting wave energy through the pores. In the regime when all length scales are deeply subwavelength we obtain the effective plate equation  \cite[\S 5.4.2]{howe1998acoustics}
\begin{equation}
\label{eq:poropleq}
  \overline{D} \Delta_\parallel^2  w_b(x,y;t)  + \overline{\rho} h_\rmP \, \partial_t^2  w_b(x,y;t) = -q,
\end{equation} 
where $\overline{D}$ and $\overline{\rho}$  are the effective stiffness and density, 
respectively, and take the form
\begin{equation}
    \overline{D} = \left(1 - \frac{2 f \nu}{1-\nu} \right)D,\quad \mbox{ and } \quad \overline{\rho} = (1-f) \rho_\rmP, 
\end{equation}
in the small porosity limit, where $f$ denotes the porosity.
Due to the presence of the narrow fluid channels, the kinematic condition \eqref{eq:kin_elpl} is extended by Darcy's law and takes the form
\begin{equation}
\label{eq:newkincond}
\partial_z \Phi(x,y,-d;t) = \partial_t w_b(x,y;t)  - K_\rmP/(\mu_\rmF h_\rmP)  P\big|^+_-,
\end{equation}  
where $K_\rmP$ is the permeability of the plate (with $\mathrm{Re} ( K_\rmP)>0$  related to dissipative effects and $\mathrm{Im} (K_\rmP)$ related to inertial effects in the flow) and $\mu_\rmF$ is the dynamic viscosity. The pressure jump $P\big|_{-}^{+}$ is given by the linearised Bernoulli form \eqref{eq:forcingq}. We refer to Tuck  \cite[\S~V.D]{tuck75} for further details and a detailed discussion of this model.  Substituting the new kinematic condition \eqref{eq:newkincond} into the effective plate equation \eqref{eq:poropleq} and removing $y$-dependence as before, we obtain the poroelastic plate equation
\begin{equation}
\label{eq:plateeqnewporo}
(\partial_x^4 - \overline{\mu}^4) (\partial_z \phi(x,z) + \overline{\tau}  \phi(x,z)\big|_{-}^{+}) + \overline{\beta}  \phi(x,z)\big|_{-}^{+} = 0, \quad \mbox{for } z=-d\mbox{ and }x >0, 
\end{equation}
where $\overline{\mu}^4 =  \overline{\rho} h_\rmP \omega^2/  \overline{D}$, $ \overline{\tau} =   \rmi \omega \rho_\rmF K_\rmP/(\mu_\rmF h_\rmP)$, and   $\overline{\beta} =  \omega^2 \rho_\rmF/ \overline{D}$   are known constants. Hence, we solve the system \eqref{eq:2dlaplacefluid}--\eqref{eq:asyfluidflow} once more,   replacing the elastic plate equation \eqref{eq:plate1} with  \eqref{eq:plateeqnewporo}. 
In the limit of zero permeability $K_\rmP \rightarrow 0$ we recover the system specified earlier. 
 As discussed in Jaworski \& Peake \cite{jaworski2013aerodynamic},  the effective plate equation above is a lowest-order approximation to a fully coupled poroelastic plate system, such as those described in Theodorakopoulos \& Beskos  \cite{theodorakopoulos1994flexural}.

We now proceed as before and renormalise the modified incident potential as
\begin{equation}
\label{eq:pwansantz2poro}
\phi^\mathrm{inc}(x,z)= \frac{-\cosh(\ell(z+h))}{\ell (\ell^4 - \overline{\mu}^4)  \sinh(\ell c) } \rme^{\rmi \ell x} ,   
\end{equation}
admitting the poroelastic plate equation
\begin{equation}
\label{eq:plate2poro}
(\partial_x^4 - \overline{\mu}^4) (\partial_z \phi^\rms(x,z) + \overline{\tau}  \phi^\rms(x,z)\big|_{-}^{+}) + \overline{\beta}  \phi^\rms(x,z)\big|_{-}^{+} = \rme^{\rmi\ell x}, \quad \mbox{for } z=-d\mbox{ and }x >0. 
\end{equation}
The   Fourier transform of the   poroelastic plate operator gives the Wiener--Hopf equation
\begin{equation}
\label{eq:wheqnporo}
\overline{K}(s) A^+(s) = \overline{P}(s) +(s^4 - \overline{\mu}^4) B^-(s)    + \frac{\rmi}{(s+\ell)_+},
\end{equation}
which is analogous to \eqref{eq:wheqn}, where
\begin{equation}
\label{eq:Ksporo}
\overline{K}(s) = s (s^4 - \overline{\mu}^4)   \left[\frac{ \sinh(sc) \eta(s) }{   \sinh(sc) \xi(s) - \cosh(sc) \eta(s) }\right] + \overline{\tau} (s^4 -  \overline{\mu}^4)   + \overline{\beta}.
\end{equation}
The decomposition of $\overline{K}(s) = \overline{L}(s) \overline{Q}(s)$ proceeds identically to before with  
\begin{subequations}
\begin{align}
\overline{L}(s) &= -\frac{1}{2}s (s^4 - \overline{\mu}^4) \frac{\cosh(\pi s)}{\sinh(\pi s)}, \\
\overline{Q}(s) &=   2 \tanh(\pi s)  \left[\frac{\sinh(cs) \eta(s) }{ \alpha \cosh(s h) - s \sinh(s h)   } - \frac{\overline{\beta}}{s(s^4-\overline{\mu}^4)} - \frac{\overline{\tau}}{s} \right]  .    
\end{align}
\end{subequations}
As before, we obtain the system 
 \begin{subequations}
 \begin{align}
\label{eq:WHalmost1poro}
 \overline{K}^+(s) A^+(s) - \frac{\rmi}{(s+\ell)_+} \frac{1}{\overline{K}^-(-\ell)}&= \overline{J}(s), \\
\label{eq:WHalmost2poro}
 \frac{ \overline{P}(s)}{\overline{K}^-(s)} +(s^4 - \overline{\mu}^4)   \frac{B^-(s)}{\overline{K}^-(s)}  + \frac{\rmi}{(s+\ell)_+}\left(     \frac{1}{\overline{K}^-(s)}   -  \frac{1}{\overline{K}^-(-\ell)} \right)   &= \overline{J}(s),
\end{align}
\end{subequations}
where $\overline{J}(s) = \overline{p}s + \overline{q}$ and both $\overline{p}$ and $\overline{q}$ are unknown.  To determine the polynomial 
 \begin{equation}
\overline{P}(s) =   \rmi (   \partial_z \phi^\rms_0 + \overline{\tau} \phi^\rms_0     \big|_{-}^{+})s^3  - (\partial_x \partial_z \phi^\rms_0 + \overline{\tau} \partial_x \phi^\rms_0     \big|_{-}^{+})  s^2  - \rmi (  \partial_x^2 \partial_z \phi^\rms_0 + \overline{\tau} \partial_x^2 \phi^\rms_0     \big|_{-}^{+}) s + (\partial_x^3 \partial_z \phi^\rms_0 + \overline{\tau}\partial_x^3 \phi^\rms_0     \big|_{-}^{+})  ,
\end{equation}
where subscript zero notation denotes, for example,
\begin{equation}
  \partial_z \phi^\rms_0 + \overline{\tau} \phi^\rms_0     \big|_{-}^{+} = \lim_{x\rightarrow 0} \left\{\partial_z \phi^\rms(x,-d)+ \overline{\tau} \phi^\rms(x,-d)     \big|_{-}^{+} \right\},   
\end{equation}
  we substitute   the kinematic condition \eqref{eq:newkincond} into the boundary conditions for the plate  to obtain 
\begin{subequations}
\label{eq:allbcsplateporo}
\begin{align}
\mbox{clamped edge}: \quad & \partial_z \phi^\rms_0 + \overline{\tau} \phi^\rms_0     \big|_{-}^{+}=  \frac{1}{\ell^4-\overline{\mu}^4}, \quad \partial_x\partial_z \phi^\rms_0 + \overline{\tau} \partial_x \phi^\rms_0     \big|_{-}^{+}=  \frac{\rmi \ell}{\ell^4-\overline{\mu}^4}, \\
\mbox{simply supported edge}: \quad & \partial_z \phi^\rms_0 + \overline{\tau} \phi^\rms_0     \big|_{-}^{+}=  \frac{1}{\ell^4-\overline{\mu}^4}, \quad \partial_x^2\partial_z \phi^\rms_0 + \overline{\tau}\partial_x^2 \phi^\rms_0     \big|_{-}^{+}= -\frac{\ell^2}{\ell^4-\overline{\mu}^4}, \\
\mbox{free edge}: \quad & \partial_x^2\partial_z \phi^\rms_0 + \overline{\tau}\partial_x^2 \phi^\rms_0     \big|_{-}^{+}= - \frac{\ell^2}{\ell^4-\overline{\mu}^4}, \quad \partial_x^3\partial_z \phi^\rms_0 + \overline{\tau}\partial_x^3 \phi^\rms_0     \big|_{-}^{+}=  -\frac{\rmi \ell^3}{\ell^4-\overline{\mu}^4},
\end{align}
\end{subequations}
which are analogous to the conditions for an elastic plate \eqref{eq:allbcsplate}. At this stage, there are  four unknown parameters from $\overline{P}(s)$ and two unknown parameters from $\overline{J}(s)$. Two conditions are given by the choice of boundary condition at the plate edge, two conditions are obtained by requiring analyticity in the lower-half plane for  \eqref{eq:WHalmost2poro}, i.e., that
\begin{equation}
\label{eq:condfornonsing1poro}
\left[ \overline{p}s + \overline{q} -  \frac{\rmi}{(s+\ell)_+}\left(     \frac{1}{\overline{K}^-(s)}   -  \frac{1}{\overline{K}^-(-\ell)} \right) -  \frac{ \overline{P}(s)}{\overline{K}^-(s)}  \right] = 0,
\end{equation}
when $s= -\overline{\mu}$ and $s=-\rmi \overline{\mu}$. Two further conditions are obtained by taking the half-range Fourier transform of the forced poroelastic plate equation \eqref{eq:plate2poro} which requires that
\begin{equation}
 \left[ \overline{P}(s) + \frac{\rmi}{(s+\ell)_+} -\frac{1}{\overline{K}^-(-\ell) \, \overline{K}^+(s)}  \left\{\overline{\tau}(s^4-\overline{\mu}^4) +\overline{\beta} \right\} \left( \overline{p}s + \overline{q} +  \frac{\rmi}{(s+\ell)_+} \right)\right] = 0   ,
\end{equation}
when $s= \overline{\mu}$ and $s=\rmi \overline{\mu}$. With these six parameters now determined, the Wiener--Hopf solution for a submerged poroelastic plate yields the two functions $A^+(s)$ and $B^-(s)$ explicitly.
\subsection{Solution representation}
 Having determined the necessary polynomials above, we return to \eqref{eq:WHalmost1poro} to write
 \begin{equation}
A^+(s)=   \frac{1}{\overline{K}^+(s) }\left(    \overline{p}s + \overline{q} + \frac{\rmi}{(s+\ell)_+} \frac{1}{\overline{K}^-(-\ell)} \right).
  \end{equation}
Using the relationship between $H^\rmL(s)$ and $A^+(s)$ in \eqref{eq:fouriertransApD}, the relationship between $G^\rmU(s)$ and $H^\rmL(s)$ in \eqref{eq:CsvsDs}, the general solution for $\Psi(s,z)$ in \eqref{eq:psipartition} is now completely prescribed for a poroelastic plate. The total field for $x<0$ is defined as before but is now evaluated as
  \begin{subequations}
\begin{equation}
\mathcal{A}\phi(x,z)  =\phi^\mathrm{inc}(x,z)+  
   \sum_j \rme^{-\rmi s_j^\mathcal{R} x} 
\begin{cases} 
       \overline{\tau}^\rmU(s_j^\mathcal{R})       \left[  \cosh(s_j^\mathcal{R} z) +\dfrac{\alpha}{s_j^\mathcal{R}} \sinh(s_j^\mathcal{R} z) \right]  &  \mbox{for }  -d<z \leq0 \\
\\
   \overline{\tau}^\rmL(s_j^\mathcal{R})      \cosh(s_j^\mathcal{R}(z+h))      & \mbox{for }  -h\leq z<-d  
   \end{cases},  
\end{equation}
where
\begin{align}
\overline{\tau}^\rmU(s) &=  \frac{\rmi}{\overline{K}^+(s) }\left(    \overline{p}s + \overline{q} + \frac{\rmi}{(s+\ell)_+} \frac{1}{\overline{K}^-(-\ell)} \right)  \frac{ s \sinh(s c) }{ (1-\alpha s) \sinh(s h) +  s^2 \cosh(s h)} , \\
\overline{\tau}^\rmL(s) &= \frac{\rmi}{\overline{K}^+(s) }\left(    \overline{p}s + \overline{q} + \frac{\rmi}{(s+\ell)_+} \frac{1}{\overline{K}^-(-\ell)} \right)  \frac{ \eta(s)}{ (1-\alpha s ) \sinh(s h) + s^2 \cosh(s h)} ,
\end{align}
\end{subequations}
and   $s_j^\mathcal{R}$ denote solutions to $  s  \sinh(sh) -\alpha \cosh(sh) =0$ in $U_+$, which are all simple. Note those zeros are identical to that found previously, but that the field amplitudes now differ. Since there is no discontinuity for $x<0$ the above expression  can be simplified to 
\begin{equation}
\mathcal{A}\phi(x,z)  =\phi^\mathrm{inc}(x,z)+  
   \sum_j \rme^{-\rmi s_j^\mathcal{R} x} \,
   \overline{\tau}^\rmL(s_j^\mathcal{R})      \cosh(s_j^\mathcal{R}(z+h))    \,\,  \mbox{for }  -h\leq z<0.  
\end{equation}

Likewise   the total field for $x>0$  is evaluated via
\begin{align}
\mathcal{A}\phi(x,z) =  \sum_j   \rme^{\rmi  \overline{s}_j^\mathcal{T} x}   \begin{cases} 
          \overline{\kappa}^\rmU( -\overline{s}_j^\mathcal{T}) \left[  \cosh( \overline{s}_j^\mathcal{T} z) +\dfrac{\alpha}{ \overline{s}_j^\mathcal{T}} \sinh( \overline{s}_j^\mathcal{T} z) \right]  &  \mbox{for }  -d<z \leq0 \\
\\
      \overline{\kappa}^\rmL( -\overline{s}_j^\mathcal{T}) \cosh( \overline{s}_j^\mathcal{T}(z+h))      & \mbox{for }  -h\leq z<-d  
   \end{cases},   
\end{align}
where
\begin{align}
\overline{\kappa}^\rmU(s) &=   -\rmi  \overline{K}^-(s) \left(    (\overline{p}s + \overline{q})(s+\ell)_+  + \frac{\rmi}{\overline{K}^-(-\ell)} \right) \frac{s \sinh(s c) }{\partial_s \overline{y}(s)}      , \\
\overline{\kappa}^\rmL(s) &=  -\rmi  \overline{K}^-(s) \left(    (\overline{p}s + \overline{q})(s+\ell)_+  + \frac{\rmi}{\overline{K}^-(-\ell)} \right)  \frac{\eta(s)}{\partial_s \overline{y}(s)} ,
\end{align}
and 
\begin{equation}
\overline{y}(s) =\left[s(s^4-\overline{\mu}^4) \sinh(sc) \eta(s) + \left\{\overline{\beta} + \overline{\tau}(s^4 - \overline{\mu}^4)\right\}\left\{s \sinh(sh)-\alpha \cosh(sh)\right\} \right] (s+\ell)_+.
\end{equation}
In the above, $ s=-\overline{s}_j^\mathcal{T}$ are solutions to $\overline{y}(s)  = 0$ in $U_-$, excluding $\overline{s}_j^\mathcal{T} = \ell$, and   are all simple.

\subsection{Conservation of energy}

Recalling the earlier result \eqref{eq:ebalplate}, we remark that  the final integral describing the work done by the fluid on the plate no longer takes the form given in \eqref{eq:elplatecontribut} earlier, but must instead be evaluated numerically in the limit as $x_\rmW \rightarrow \infty$, i.e., as
\begin{equation}
\mathrm{Im} \left\{ \int_{0}^{\infty} (\phi(x,-d)\big|^+_-)^\ast \partial_z \phi(x,-d) \, \rmd x  \right\},
\end{equation}
so that the total dissipative loss is captured.

\subsection{Numerical examples}
In Figure~\ref{time_dependent_clamped_porous} (and the movie file in the supplementary material) we show the effect of porosity for the configuration considered in Figure~\ref{time_dependent_clamped}. We set the non-dimensional porosity to be $\tau = 0.01$. The transmitted waves can be seen to decay as they propagate and we can see that the lower wavenumber mode (which has longer wavelength) has significantly less porosity-driven decay due to its closeness to the real line. 
     \begin{figure}
\begin{center}
\includegraphics[scale = 0.4]{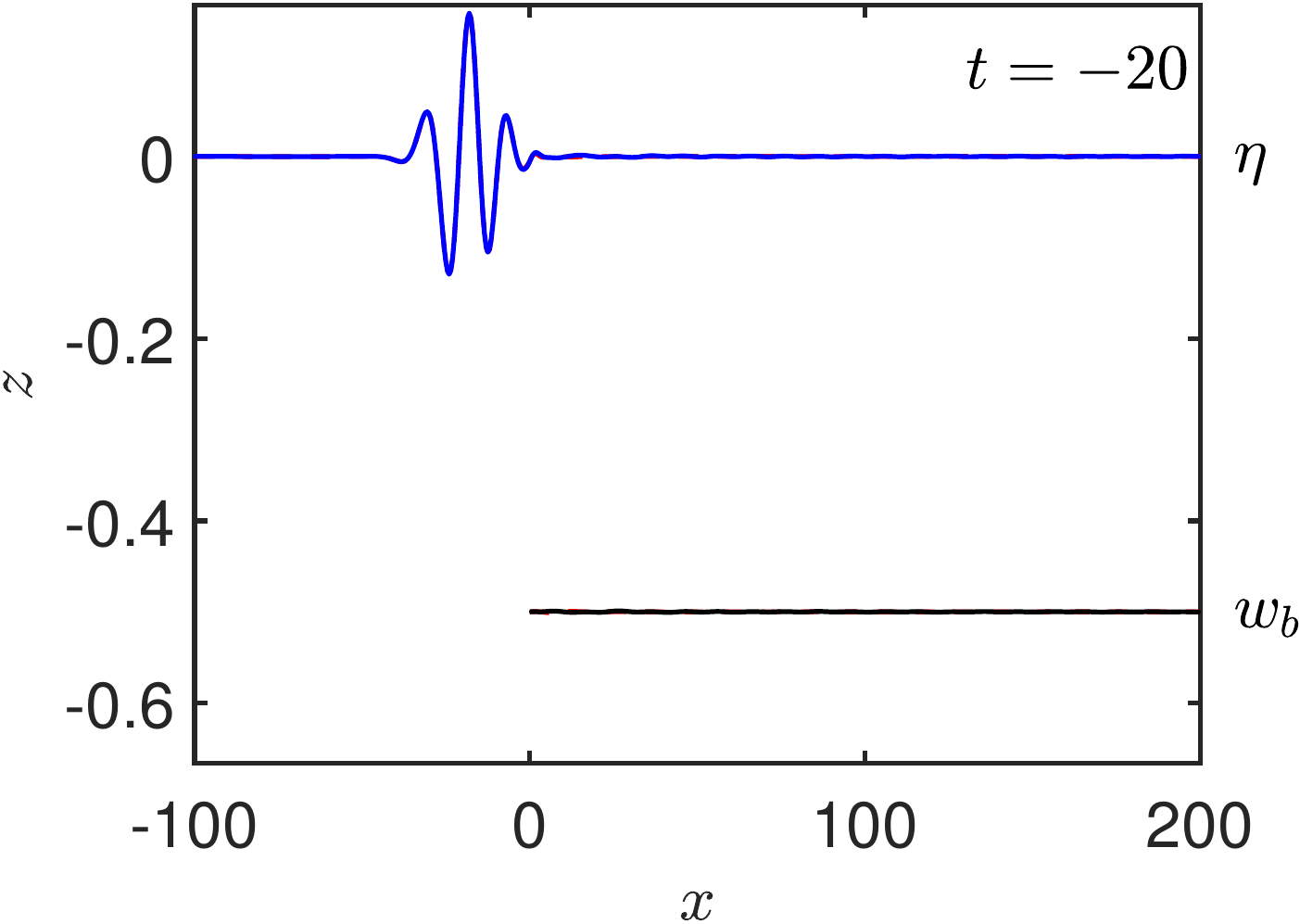}
\hspace{0.5cm}
\includegraphics[scale = 0.4]{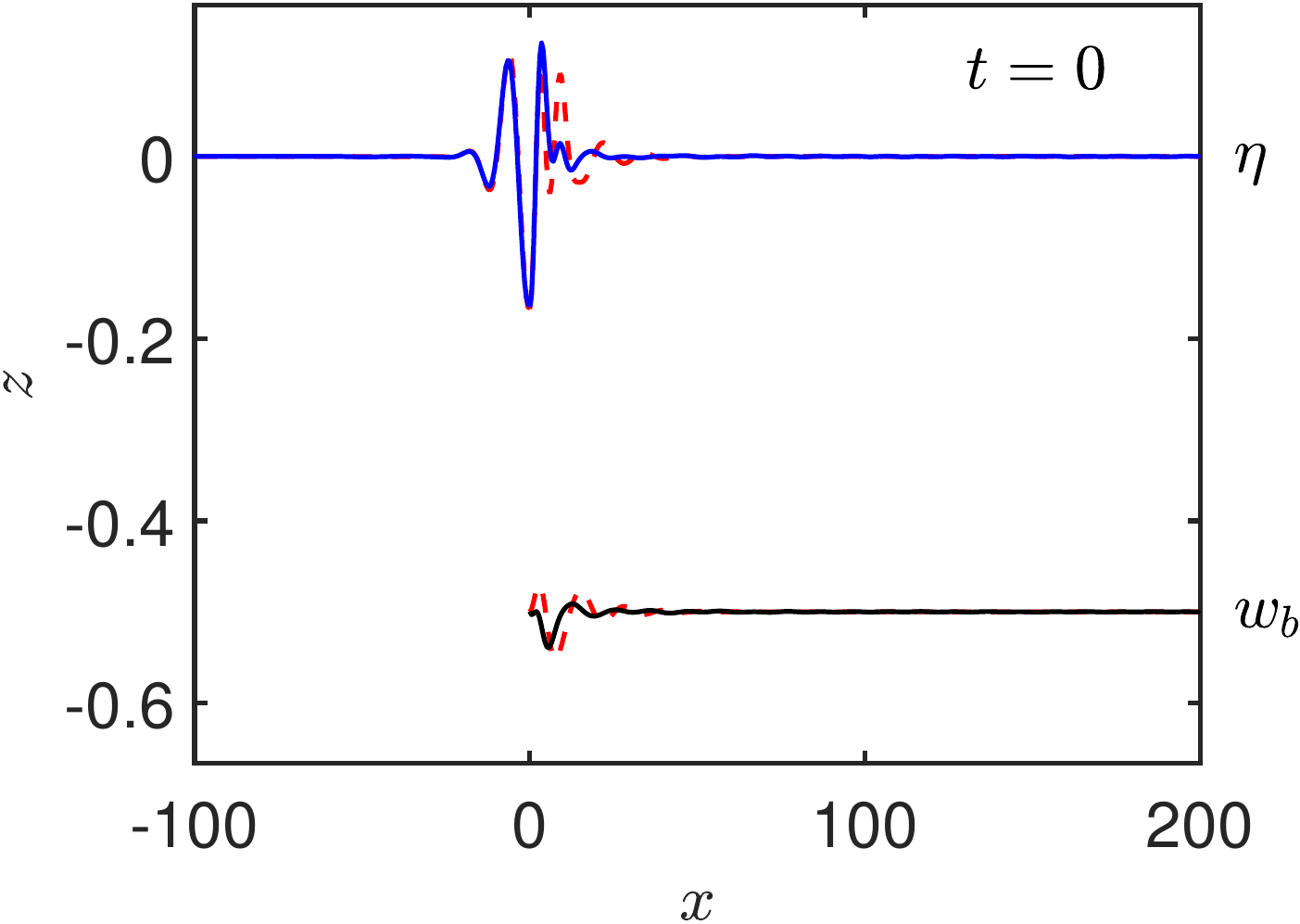}\\
\includegraphics[scale = 0.4]{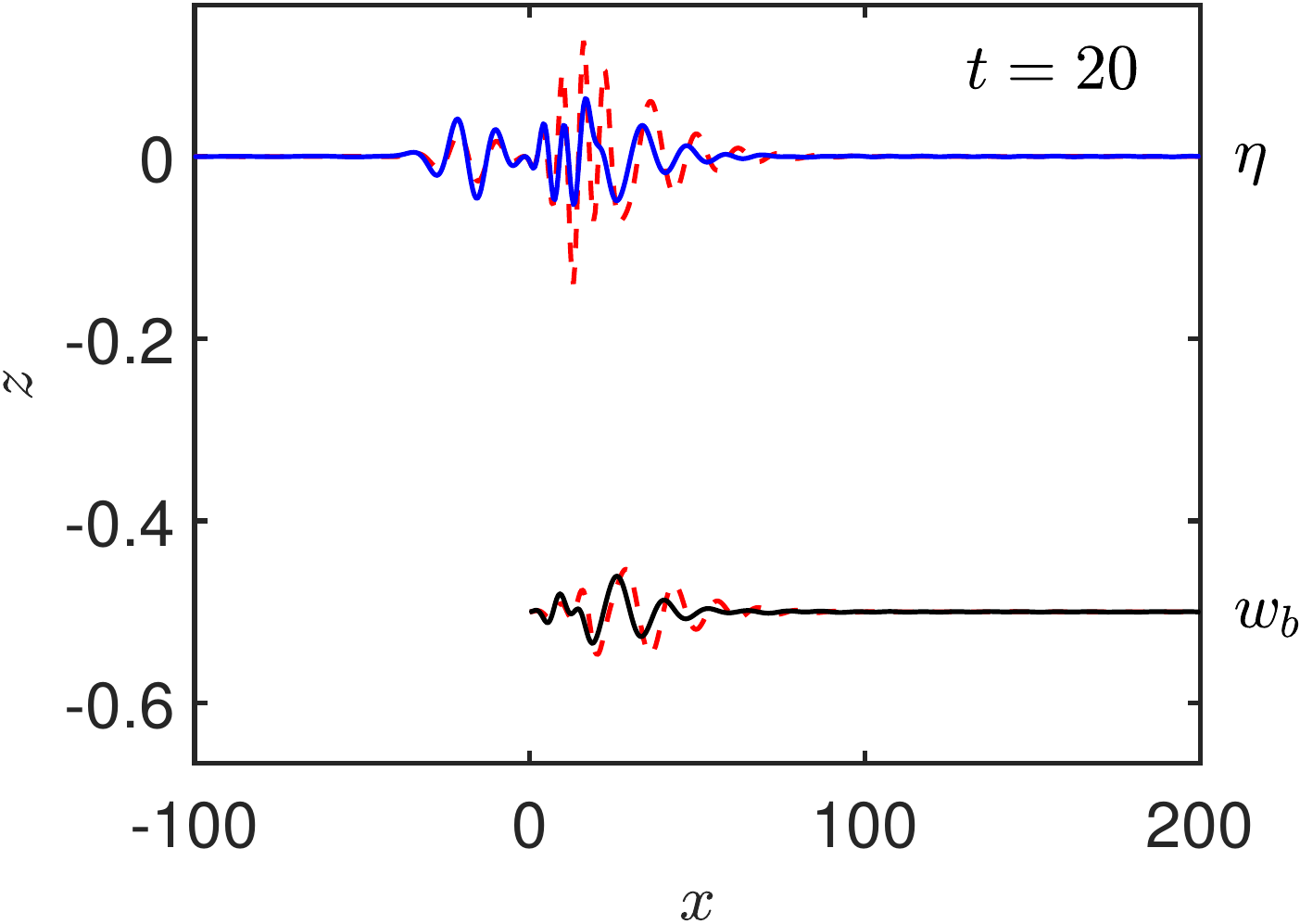}
\hspace{0.5cm}
\includegraphics[scale = 0.4]{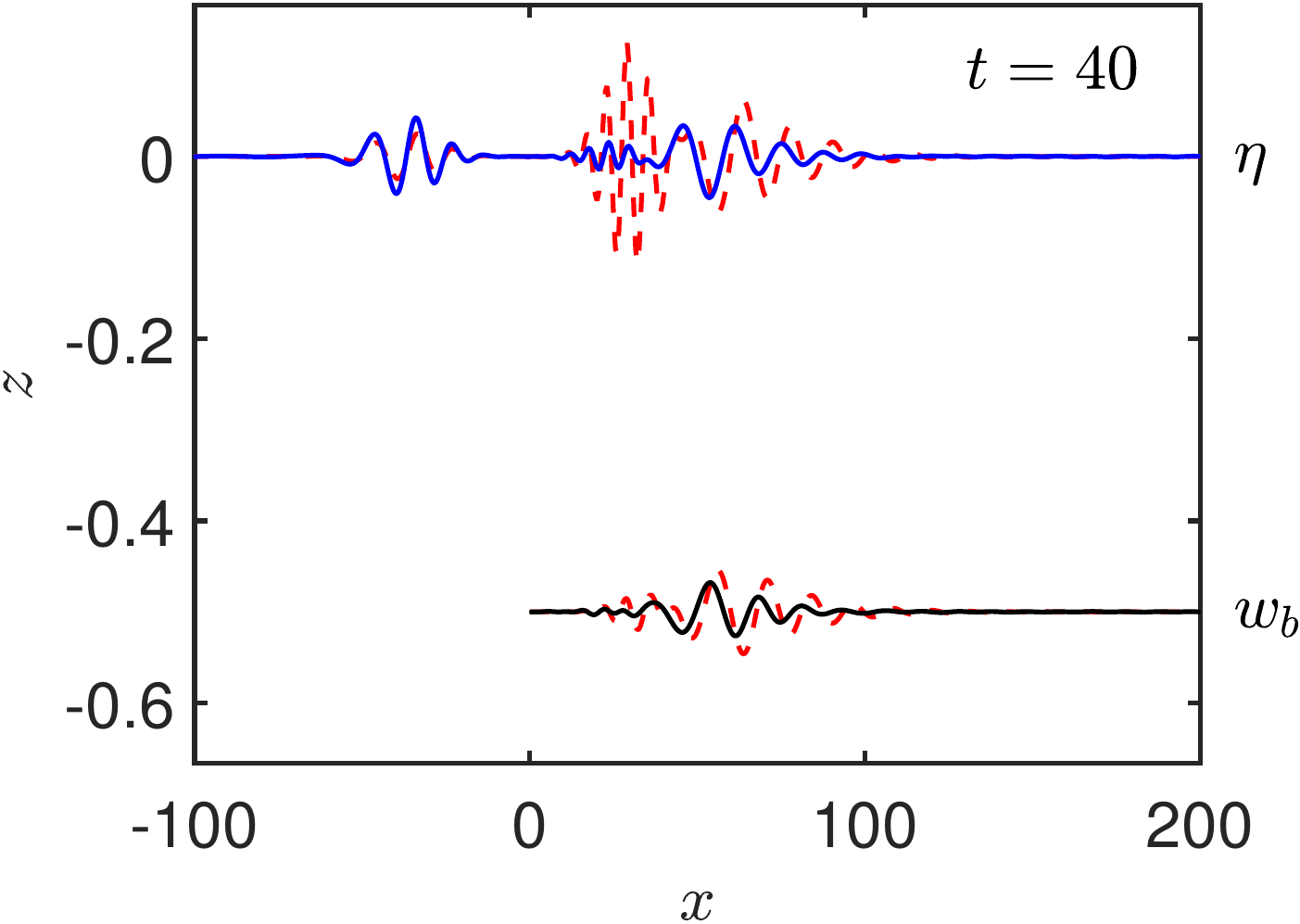}\\
\includegraphics[scale = 0.4]{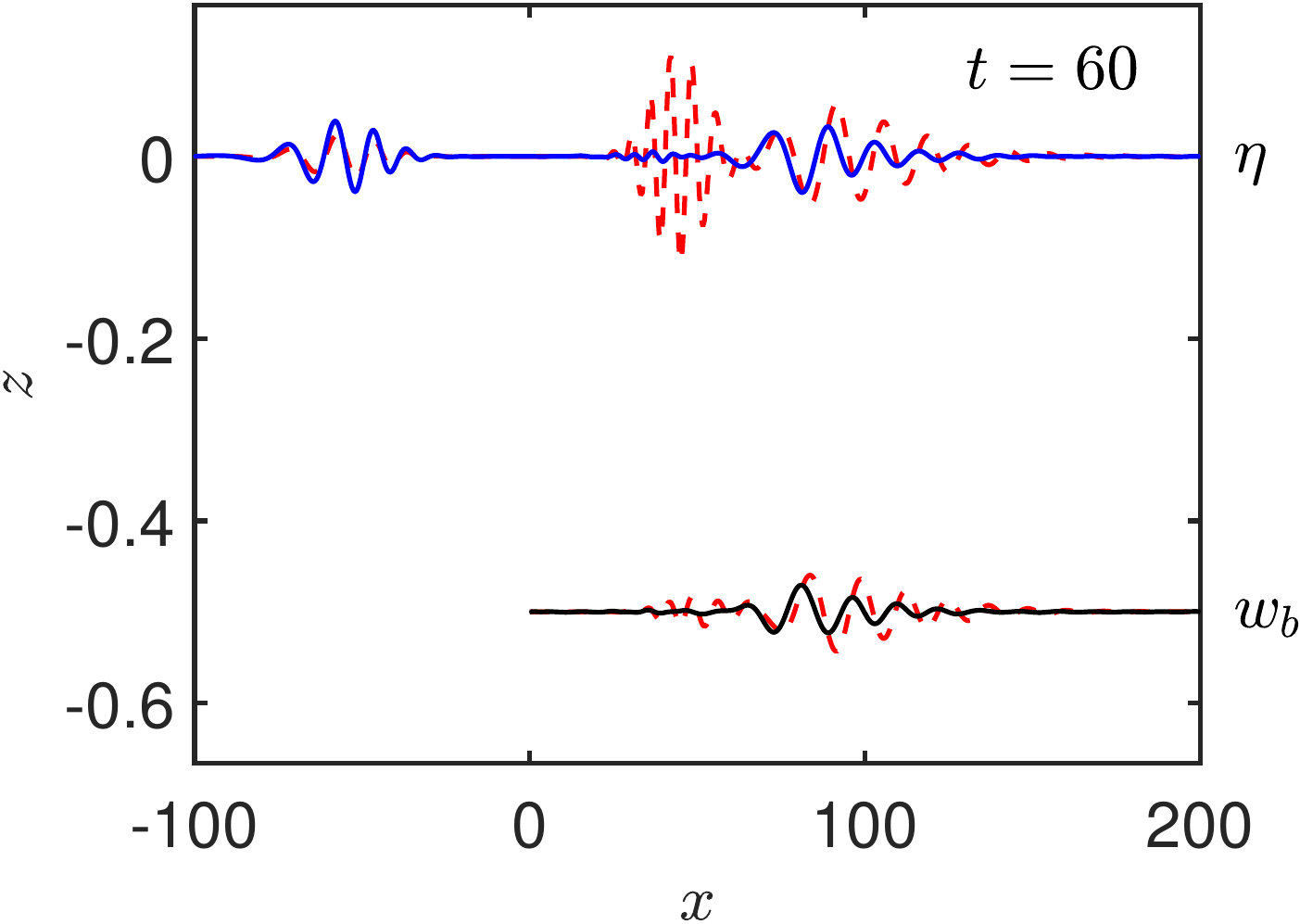}
\hspace{0.5cm}
\includegraphics[scale = 0.4]{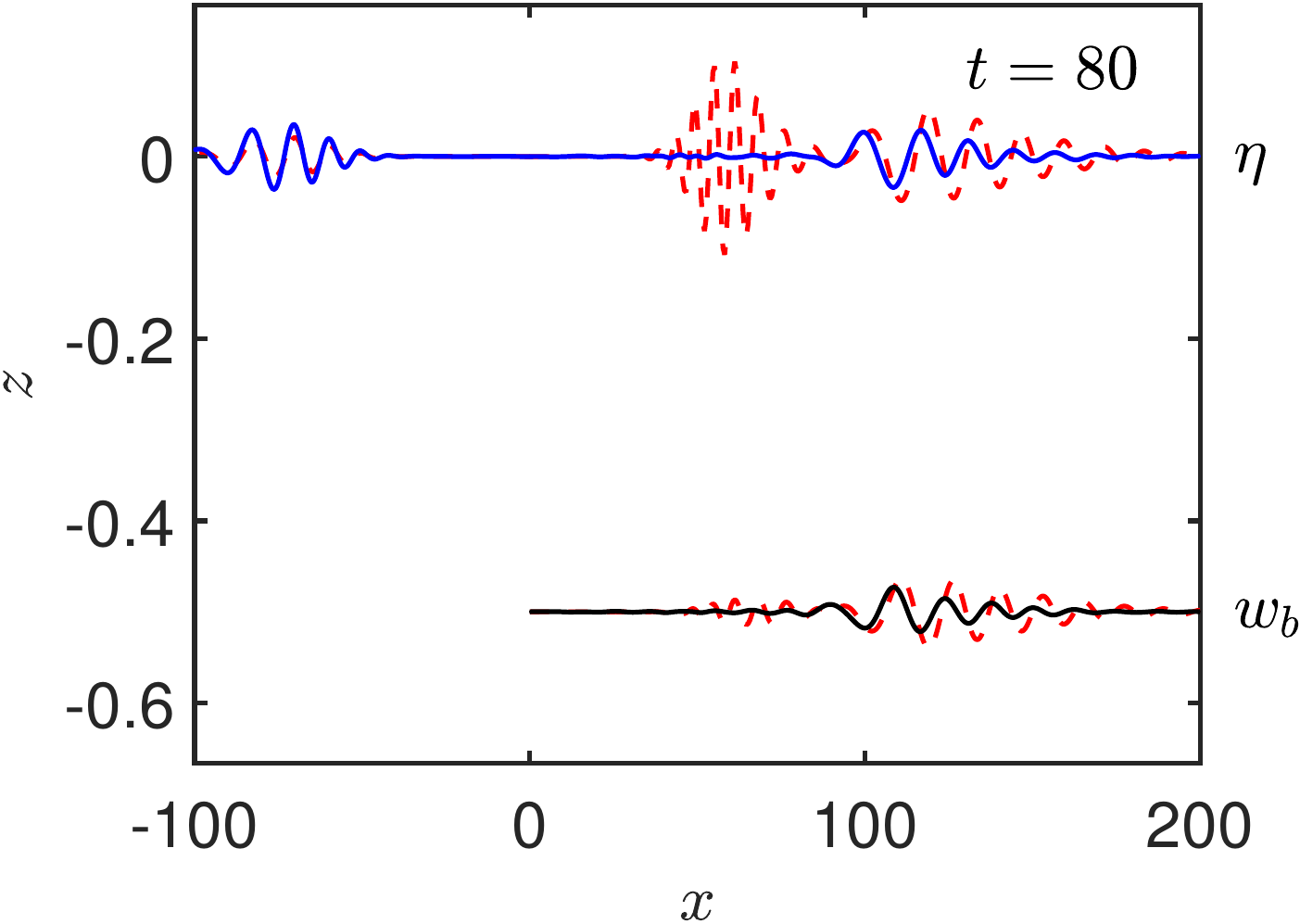}
\end{center}
\caption{\small As in Figure~\ref{time_dependent_clamped} except with porosity $\overline{\tau} = 0.01$. Also superposed is the solution without porosity (red dashed lines).The full animation can be found in movie 3 in the Supplementary Material.}%
\label{time_dependent_clamped_porous}
\end{figure}

\clearpage
\section{Concluding remarks}\label{sec:concl_rem}

We have presented a convenient Wiener--Hopf solution method for a submerged elastic plate which we have extended to the case of a submerged porous plate.  In contrast to earlier works in this field, the Wiener--Hopf equation was derived in a direct manner, and the splitting was accomplished by the use of a Cauchy-type integral. The solution method was illustrated by computing the response of a submerged plate to incident Gaussian wave pulses.  We showed that the effect of porosity is to damp propagating waves and that the short wavelength transmitted wave in the plate--fluid system is damped significantly more than the long-wavelength transmitted wave.

\section*{Acknowledgements}
All authors would like to thank the Isaac Newton Institute for
Mathematical Sciences, Cambridge, for its support and hospitality during the
programme "Bringing pure and applied analysis together via the
Wiener--Hopf technique, its generalisations and applications" where  
work on this paper was undertaken. This work was supported by EPSRC
grant no EP/R014604/1.

\bibliographystyle{RS.bst}

\end{document}